\documentclass{IEEEoj}
\usepackage{cite}
\usepackage{amssymb,amsfonts} 
\usepackage{graphicx,color}
\usepackage{textcomp}
\usepackage{multirow}
\usepackage{soul}
\usepackage{subcaption,balance}
\usepackage{comment}
\usepackage{makecell}
\usepackage{algorithm}
\usepackage{algorithmicx}
\usepackage{algpseudocode} 
\usepackage{amsmath}
\usepackage{multicol}
\usepackage[table]{xcolor}
\makeatletter 

\allowdisplaybreaks[4]

\newtheorem{theorem}{Theorem}

\newtheorem{lemma}{Lemma}

\newtheorem{proposition}{Proposition}

\newtheorem{corollary}{Corollary}

\newtheorem{property}{Property}

\newtheorem{remark}{Remark}

\newtheorem{claim}{Claim}

\makeatletter
\renewcommand{\maketag@@@}[1]{\hbox{\m@th\normalsize\normalfont#1}}%
\makeatother

\def\BibTeX{{\rm B\kern-.05em{\sc i\kern-.025em b}\kern-.08em
    T\kern-.1667em\lower.7ex\hbox{E}\kern-.125emX}}
\AtBeginDocument{\definecolor{ojcolor}{cmyk}{0.93,0.59,0.15,0.02}}
\def\OJlogo{\vspace{-4pt}\hskip-4pt\includegraphics[height=18pt]{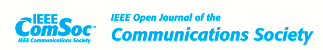}}
\begin{document}
		\makeatletter % 重点代码区开始 
	\let\myorg@bibitem\bibitem
	\def\bibitem#1#2\par{%
		\@ifundefined{bibitem@#1}{%
			\myorg@bibitem{#1}#2\par
		}{%
			\begingroup
			\color{\csname bibitem@#1\endcsname}%
			\myorg@bibitem{#1}#2\par
			\endgroup
		}%
	}
	\makeatother % 重点代码区结束 
	
% \receiveddate{01 Apr, 2025}
% \reviseddate{12 May, 2025}
% \accepteddate{19 Jun, 2025}
% \publisheddate{XX Month, XXXX}
% \currentdate{XX Month, XXXX}
% \doiinfo{OJCOMS.2024.011100}

\title{Linear and Numerical SDoF Bounds of Active RIS-Assisted MIMO Wiretap Interference Channel}

\author{\title{\relax L}Linfan Su\IEEEauthorrefmark{1}, Yuhang Miao\IEEEauthorrefmark{2},
Yuxuan Song\IEEEauthorrefmark{2},    Shuo Zheng\IEEEauthorrefmark{3}, \\ Tong Zhang\IEEEauthorrefmark{2,5},  Yinfei Xu\IEEEauthorrefmark{4}, Shuai Wang\IEEEauthorrefmark{6}, Na Li\IEEEauthorrefmark{1} }
\affil{National Engineering Laboratory for Mobile Network Security, Beijing University of Posts and Telecommunications, Beijing, China}
\affil{Institute of Intelligent Ocean Engineering, Harbin Institute of Technology (Shenzhen), Shenzhen, China}
\affil{Department of Electrical and Electronic Engineering, The Hong Kong Polytechnic University, Hong Kong SAR, China}
\affil{School of Information Science and Engineering, Southeast University, Nanjing, China}
\affil{National Mobile Communications Research Laboratory, Southeast University, China}
\affil{Shenzhen Institutes of Advanced Technology, Chinese Academy of Sciences, Shenzhen, China}
\corresp{CORRESPONDING AUTHOR: Tong Zhang (e-mail: tongzhang@hit.edu.cn), Na Li  (e-mail: lina\_lena@bupt.edu.cn).}
% \authornote{This work was supported in part by the National Natural Science Foundation of China under Grant No.62371062, No.62401229, No.62371119 and No.62371444; in part by the Fundamental Research Funds for the Central Universities under Grant No.2242022k60006; in part by the Shenzhen Science and Technology Program under Grant No.RCYX20231211090206005 and
% No.JCYJ20241202124934046; in part by the Guangdong Basic and Applied Basic Research Project under Grant No.2023A1515110477; in part by the open research fund of National Mobile Communications Research Laboratory, Southeast University under Grant No.2025D07.}
\markboth{Preparation of Papers for IEEE OPEN JOURNALS}{Author \textit{et al.}}

\begin{abstract}
The multiple-input multiple-output (MIMO) wiretap interference channel (IC) serves as a canonical model for information-theoretic security, where a multiple-antenna eavesdropper attempts to intercept communications in a two-user MIMO IC system. The secure degrees-of-freedom (SDoF) of an active reconfigurable intelligent surface (RIS)-assisted MIMO wiretap IC is with practical interests but remains unexplored. In this paper, we establish both sum-SDoF lower and upper bounds through linear beamforming conditions and numerical methods. Specifically, our proposed lower bound is derived from transmission scheme design and corresponding solutions to the sum-SDoF maximization problem, formulated by linear integer programming. The solutions to this optimization problem addresses RIS element allocation for leakage and interference cancellation. The proposed upper bound is obtained by solving a nuclear norm minimization problem, leveraging the fact that nuclear norm serves as a convex relaxation of the rank function. For symmetry antenna configurations, we derive a closed-form lower bound. Extensive numerical simulations show that our proposed lower and upper bounds coincide across many antenna configurations, and our proposed lower bound outperforms the existing benchmark.
\end{abstract}

\begin{IEEEkeywords}
Active RIS, linear beamforming, sum-SDoF, MIMO wiretap IC, 
\end{IEEEkeywords}

\maketitle

\section{INTRODUCTION}
\IEEEPARstart{M}{ultiple}-input multiple-output (MIMO) technology stands as a cornerstone of modern wireless communication systems, with its significance poised to expand further in the forthcoming 6G networks. MIMO systems leverage multiple antennas at both transmitter and receiver ends to substantially enhance wireless network capacity, spectral efficiency, and coverage. Through the exploitation of multiplexing techniques, MIMO enables the concurrent transmission of multiple data streams across identical frequency bands, thereby boosting overall data throughput without demanding additional bandwidth resources\cite{he2021cell,Larsson,Cadambe}.
Nevertheless, the inherent openness of wireless communication leaves it vulnerable to eavesdropping, as radio signals can be intercepted by unauthorized parties within transmission range. To address this vulnerability, physical layer security has emerged as a promising complement to traditional upper layer security techniques, exploiting the fundamental characteristics of wireless channels to establish secure communication \cite{schaefer2017physical,Dong,Mukherjee}. Secrecy capacity is the key measure of physical layer security, defining the maximum rate at which information can be reliably transmitted to legitimate receivers while ensuring eavesdroppers gain virtually no knowledge of the message content. This fundamental metric, calculated as the difference between legitimate and eavesdropper channel capacities, serves as the fundamental benchmark for evaluating the effectiveness of physical layer security techniques\cite{liu2009note,oggier2011secrecy,Gopala}.
However, the exact secrecy capacity for most multi-user wiretap channels remains elusive, despite significant progress in coarser characterizations such as secure degrees-of-freedom (SDoF). SDoF has emerged as a valuable first-order approximation metric for secrecy capacity in the high signal-to-noise ratio (SNR) regime. Also referred to as secure multiplexing gain, SDoF represents the maximum number of orthogonal data streams that can be transmitted with security guarantees. 

% \textcolor{blue}{~\cite{zhang2021secure}}

SDoF metric has been extensively investigated across numerous studies \cite{xie2014secure,banawan2015secure,BanawanTIT,wang2019secure,zhang2021secure,Sheng,zhang2022secure}.
Numerous studies have investigated the SDoF for the two-user MIMO interference channel (IC), as this model serves as a fundamental framework for interference analysis. For the two-user MIMO IC with confidential messages (ICCM) featuring symmetry antenna configurations, the exact sum-SDoF was established in ~\cite{banawan2015secure}, with transmission schemes for arbitrary full-rank antenna configurations demonstrated through interference alignment techniques. In \cite{BanawanTIT}, the authors thoroughly examined the two-user MIMO ICCM, deriving the precise SDoF region for symmetry antenna configurations while introducing innovative transmission schemes that combine real and spatial interference alignment approaches. For time-varying MIMO ICCM scenarios, researchers developed simplified schemes that exploit channel variations to achieve secure transmission. In practical environments where channels may exhibit low-rank characteristics, the sum-SDoF of the two-user MIMO ICCM with low-rank channels was analyzed in \cite{wang2019secure}. The work in ~\cite{zhang2021secure} derived and optimized the sum-SDoF of the two-user MIMO IC with local output feedback across various antenna configurations. Beyond ICCM, researchers have explored related channel model, i.e., broadcast channel with confidential messages (BCCM). The sum-SDoF of the two-user MIMO BCCM and delayed CSIT was characterized in \cite{Sheng}. Thereafter, the sum-SDoF of the three-user MIMO BCCM with delayed CSIT was investigated in \cite{zhang2022secure}.

Recent studies have demonstrated that reconfigurable intelligent surfaces (RIS) can significantly enhance the efficiency and security of wireless communications when integrated with MIMO systems \cite{YangRIS,Sixian,HongRIS}.
From a DoF perspective, RIS-assisted MIMO channels have been extensively investigated in~\cite{seddik2022degrees,cheng2021multiplexing,cheng2023degree,bafghi2022degrees,Weihua,Feng,chae2022cooperative,zheng2023dof,OJCOMS,nafea2021secure,luo2024secure,Azari}. The existing work can be categorized into: 1) passive RIS-assisted communication \cite{seddik2022degrees,cheng2021multiplexing,cheng2023degree}; and 2) active RIS-assisted communication \cite{bafghi2022degrees,Weihua,Feng,chae2022cooperative,zheng2023dof,OJCOMS,nafea2021secure,luo2024secure,Azari}.
Related work on DoF of a passive RIS assisting communications focus on information transmission by passive RIS \cite{seddik2022degrees,cheng2021multiplexing,cheng2023degree}. 
In \cite{seddik2022degrees}, this paper explored the integration of passive RIS with non-coherent MIMO communications, where the key contribution is the derivation of the achievable DoF gained by RIS, showing that RIS can increase DoF and improve symbol error rate  performance through phase modulation. The authors of \cite{cheng2021multiplexing} and \cite{cheng2023degree} showed that passive  RIS  can significantly improve wireless communication system performance by modulating information through adjustable phase shifts, achieving a larger DoF than that without a RIS.

Active RIS integrates active amplifiers (e.g., power amplifiers) in its reflecting elements, which can compensate for path loss during signal transmission and overcome the ``multiplicative fading'' effect of traditional passive RIS \cite{Dai,ActiveFeng,ActiveFengone}.
Extensive research has been conducted on the DoF of active  RIS assistance in communication systems, with particular emphasis on interference elimination applications \cite{bafghi2022degrees,Weihua,Feng,chae2022cooperative,zheng2023dof,OJCOMS,nafea2021secure,luo2024secure,Azari}. In \cite{bafghi2022degrees}, Bafghi et al. analyzed the DoF of $K$-user IC with both active and passive RIS assistance, demonstrating that the sum DoF can approach $K$ (versus $K/2$ without RIS) when sufficient RIS elements are deployed, with active RIS achieving this benchmark using finite elements, while passive RIS requires infinite time slots and number of RIS elements. In a complementary approach, the authors of \cite{Weihua} developed a novel interference subspace alignment scheme for $K$-user MIMO IC, employing joint active and passive beamforming where passive components compress the interference subspace and active components eliminate remaining interference. Further advancing this field, the work in \cite{Feng} examined DoF and beamforming designs for active RIS-assisted $K$-user MIMO IC in rank-deficient channels, revealing significant enhancements in DoF and sum-rate performance, especially in challenging low-rank Line-of-Sight (LoS) scenarios. Furthermore, the authors in \cite{chae2022cooperative} investigated $K$-user rank-deficient MIMO IC with active RIS assistance, deriving lower and upper DoF bounds that converge under specific conditions. A notable contribution by \cite{zheng2023dof} established the achievable DoF for active RIS-assisted two-user MIMO IC with arbitrary antenna configurations, identifying specific DoF gains and characterizing optimal antenna arrangements. Additionally, the research in \cite{OJCOMS} explored how RIS can enhance the DoF of wireless multi-user X-networks, establishing that the theoretical sum-DoF becomes achievable when the number of active RIS elements exceeds a specific threshold.

Simultaneously with active RIS assistance and security constraints, active RIS assisted communication was examined from the SDoF perspective in  \cite{nafea2021secure,luo2024secure,Azari}. In \cite{nafea2021secure},  this paper investigated the SDoF of an active RIS-assisted MIMO wiretap channel, establishing both lower and upper bounds while revealing the connection between the upper bound and rank minimization problems. In \cite{luo2024secure}, the authors provided a comprehensive characterization of the sum-SDoF for an active RIS-assisted two-user MIMO broadcast channel (BC). Meanwhile, the authors of \cite{Azari} explored both the DoF and SDoF of $K$-user MIMO IC with instantaneous relays, which introduces restricted interference alignment and transmission in the null space schemes that substantially enhance both communication efficiency and security performance.

The wiretap IC is known as a fundamental and canonical model for information-theoretic security, where a multiple-antenna eavesdropper attempts to intercept the messages transmitted during IC communications \cite{Park} and \cite{Kong}. Nevertheless, the optimal transmission schemes and theoretical sum-SDoF regarding the integration of an active-RIS assistance in a two-user MIMO wiretap IC settings remains unclear. In this paper, we therefore investigate the sum-SDoF of an active RIS-assisted MIMO wiretap IC.  Our contributions are summarized as follows:
\begin{itemize}
\item \textit{Linear and Numerical Lower Bound}: We present a novel lower bound for the active RIS-assisted MIMO wiretap IC under linear beamforming constraints. Our proposed transmission scheme integrates: singular value decomposition (SVD), RIS-assisted interference and leakage elimination, zero-forcing (ZF) precoding, and interference decoding (ID). This scheme systematically combines and extends the techniques developed for low-rank MIMO IC in \cite{Krishnamurthy} and active RIS-assisted MIMO IC in \cite{zheng2023dof}. We formulate the sum-SDoF lower bound maximization as a linear integer programming problem, which can be optimally solved by GUROBI's branch-and-bound (BnB) algorithm \cite{gurobi}.

\item \textit{Linear and Numerical Upper Bound}: We propose an upper bound for the active RIS-assisted MIMO wiretap IC under linear beamforming constraints, derived through both analytical and numerical methods. Our unified framework handles two distinct cases: In no leakage links case, We transform the linear DoF upper bound for MIMO IC from~\cite{Tang} into a closed-form expression via rank function decomposition. In leakage link
residue case, we treat the channel as a MIMO wiretap channel and thus leverage results from~\cite{nafea2021secure}. After that, the key innovation lies in nuclear norm minimization as a convex relaxation of the rank function for numerical bound derivation, distinguishing our work from existing work.  	
	
	\item \textit{Closed-Form Lower Bound under Condition and Numerical Simulations}: We analyze and simulate the performance of the proposed lower and upper bounds. Furthermore, we derive a closed-form expression for the sum-SDoF lower bound under symmetry antenna configurations.  Numerical simulations  show that our proposed lower and upper bounds coincide with each other for a lots of antenna configurations. Moreover, our proposed lower bound demonstrates strict dominance over the existing bound in~\cite{zheng2023dof}, achieving 1) performance equivalence when all leakage links are completely eliminated; and 2) significant performance gains with residual leakage links.
	\end{itemize}

A comparison of this work and the existing work can be found in Table 1.  

\textit{Organization}: The system model is defined in Section-II. Our main results and discussion are given in Section-III. The proof of proposed sum-SDoF lower bound is given in Section-IV. The proof of proposed sum-SDoF upper bound is given in Section-V. We draw our conclusion in Section-VI.

\textit{Notation}:  A scalar,  a column vector, and a matrix are denoted by $a$, $\textbf{a},$ and $\textbf{A}$, respectively.  $\log$ refers to $\log_2$. $I(\cdot)$ denote  mutual information. The non-negative integer is denoted by $\mathbb{Z}^+$.  
The complex number is denoted by $\mathbb{C}$. The rank of matrix $\textbf{A}$ is denoted by $\text{rk}(\textbf{A})$. The vector of all zeros is denoted by $\textbf{0}$. The diagonal matrix is denoted by $\text{diag}\{\cdot,\cdot,\cdots,\cdot\}$. Nuclear norm is denoted by $\|\cdot\|_*$.

\begin{figure}[t]
	\centering
	\includegraphics[width=.97\linewidth]{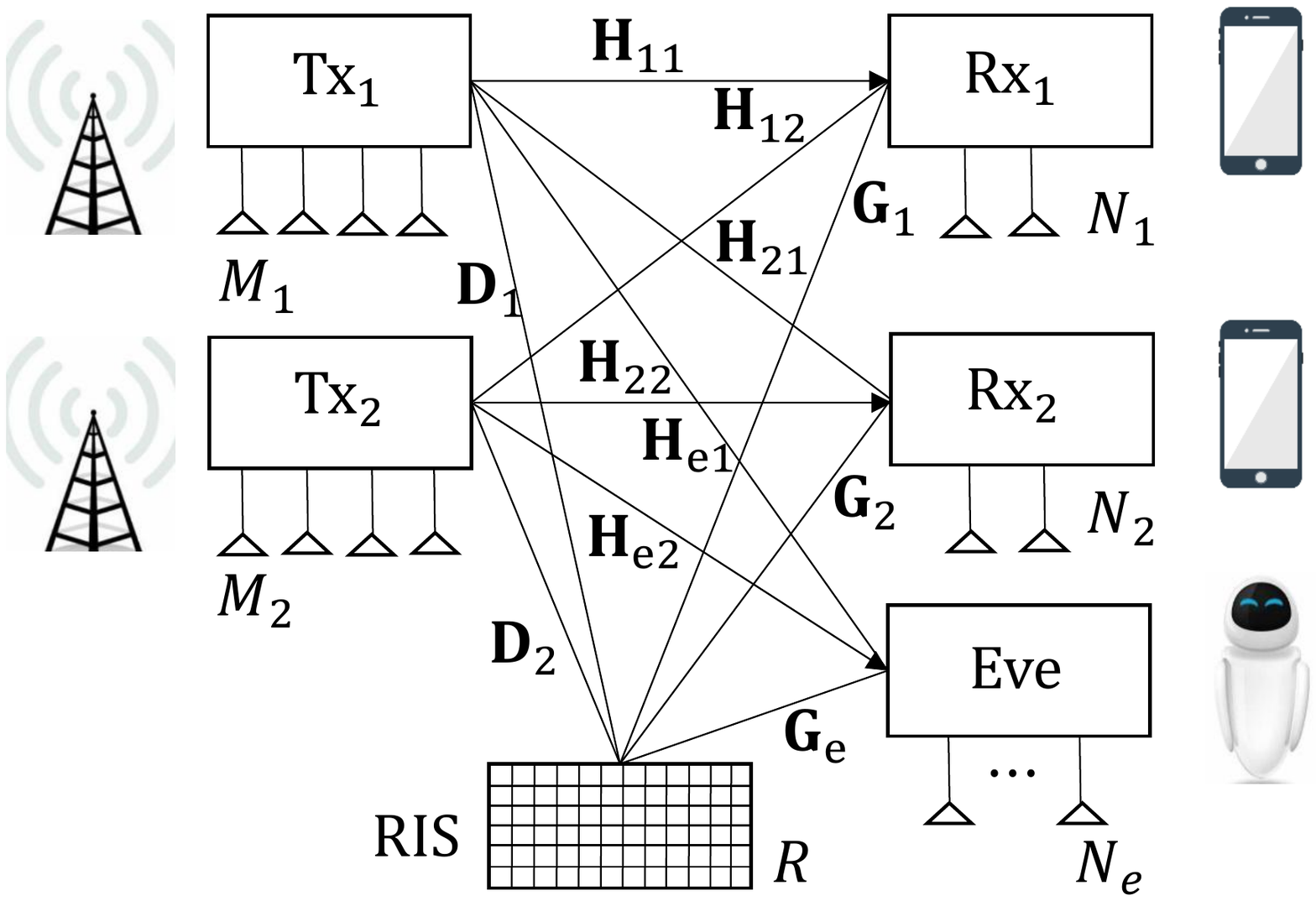}%\vspace{-0.1cm}
	\caption{Illustration of the RIS-assisted MIMO wiretap IC, where the antenna configurations are arbitrary.}
	\label{fig:figure1} 
\end{figure}	

\section{System Model}
As illustrated in Fig.~\ref{fig:figure1}, we consider an RIS-assisted two-user MIMO wiretapped IC, where the transmitters, legitimate receivers, and eavesdropper are equipped with $M_1,\,M_2$, $N_1,\,N_2$, and $N_e$ antennas, respectively. The two transmitters are denoted by $\text{Tx}_1$ and $\text{Tx}_2$, the two receivers by $\text{Rx}_1$ and $\text{Rx}_2$, and the eavesdropper as Eve. The active RIS comprises $R$ reflecting Elements, each applying both phase shift and amplitude modification to the received signals from the transmitters.

Hereafter, we restrict ourselves to linear beamforming strategies as defined
in \cite{Issa,Sridharan,Kao}, where DoF simply represents the dimension
of the linear subspace of transmitted signals and the beamforming schemes at transmitters are linear. To be specific, transmitted signals of $\text{Tx}_1$ and $\text{Tx}_2$ are denoted by $\mathbf{P}_1\mathbf{x}_1 \in \mathbb{C}^{M_1 \times 1}$ and $\mathbf{P}_2\mathbf{x}_2 \in \mathbb{C}^{M_2 \times 1}$, respectively, where  $\mathbf{x}_1\in \mathbb{C}^{\delta_1 \times 1}$ and $\mathbf{x}_2\in \mathbb{C}^{\delta_2 \times 1}$ contain the messages for $\text{Rx}_1$ and $\text{Rx}_2$, respectively, and $\mathbf{P}_1\in \mathbb{C}^{M_1 \times \delta_1}$ and $\mathbf{P}_2\in \mathbb{C}^{M_2 \times \delta_2}$ are linear transmit beamforming matrices. The received signals at Eve, $\text{Rx}_1$ and  $\text{Rx}_2$ can be expressed as follows:
    \begin{subequations}
        \begin{eqnarray}
        && \!\!\!\!\!\!\!\!\!\!\!\!\!\!\!\!\!\!\!\! \mathbf{y}_{\mathrm{e}} = \underbrace{(\mathbf{H}_{\mathrm{e}1}+\mathbf{G}_{\mathrm{e}}\mathbf{\Phi }\mathbf{D}_{1})\mathbf{P}_1\mathbf{x}_1}_{\text{leakage of } W_1} + \underbrace{(\mathbf{H}_{\mathrm{e}2}+\mathbf{G}_{\mathrm{e}}\mathbf{\Phi}\mathbf{D}_{2})\mathbf{P}_2\mathbf{x}_2}_{\text{leakage of } W_2}, \\
        && \!\!\!\!\!\!\!\!\!\!\!\!\!\!\!\!\!\!\!\! \mathbf{y}_j = \underbrace{(\mathbf{H}_{jj}+\mathbf{G}_{j}\mathbf{\Phi}\mathbf{D}_{j})\mathbf{P}_j\mathbf{x}_j}_{\text{desired signal}} + \underbrace{(\mathbf{H}_{j\overline{j}}+\mathbf{G}_{j}\mathbf{\Phi}\mathbf{D}_{\overline{j}})\mathbf{P}_{\overline{j}} \mathbf{x}_{\overline{j}}}_{\text{interference}},  
        \end{eqnarray}
    \end{subequations}
    where $j=1,2, \,\overline{j} = 3-j,$ $\mathbf{H}_{ji}\in \mathbb{C}^{N_j \times M_i},i,j=1,2$ denotes the direct channel matrix between  $\text{Tx}_i$ and  $\text{Rx}_j$, $\mathbf{H}_{\mathrm{e}i}\in \mathbb{C}^{N_e\times M_i},i=1,2$ denotes the direct channel matrix between $\text{Tx}_i$ and Eve, $\mathbf{D}_{i}\in \mathbb{C}^{R \times M_i}$ denotes the channel matrix between $\text{Tx}_i$ and the RIS, $\mathbf{G}_{j}\in \mathbb{C}^{N_j \times R}$ denotes the channel matrix between $\text{Rx}_j$ and the RIS, $\mathbf{G}_{\mathrm{e}}\in \mathbb{C} ^{N_e\times R}$ denotes the channel matrix between Eve and the RIS, $\mathbf{\Phi }=\mathrm{diag}\{\phi _1,\phi _2,\cdots ,\phi _R\}\in \mathbb{C} ^{R\times R}$ denotes the diagonal RIS reflection matrix, where the amplitude and phase of RIS element $\phi$ are obtained by $|\phi|$ and $\angle \phi$, respectively. We assume that CSI matrices are full-rank and perfectly estimated. 

	\begin{table*} 
	\caption{Comparison of This Work and Related Work on the DoF of RIS-assisted Communications}
	\label{tab:comparison}
	\centering
	\begin{tabular}{|c|c|c|c|c|c|c|c|c|c|c|c|}
		\hline
		%\rowcolor{blue}
		\textbf{Ref.} & \textbf{\makecell{Active\\RIS}} & \textbf{\makecell{Passive\\RIS}} & \textbf{$K$-User IC} & \textbf{X-Network} & \textbf{2-User IC} & \textbf{2-User BC} & \textbf{MIMO} & \textbf{\makecell{Rank\\Deficient}} & \textbf{Security} & \textbf{Lower} & \textbf{Upper} \\
		\hline
		\cite{seddik2022degrees}	&   & $\checkmark$ &  &  &  &  & $\checkmark$ &   &  & $\checkmark$ &  \\
		\hline
		\cite{cheng2021multiplexing}	&   & $\checkmark$  &  &  &   &  & $\checkmark$ &  &  & $\checkmark$ & $\checkmark$\\
		\hline
		\cite{cheng2023degree}	&   & $\checkmark$  &  &  &   &  & $\checkmark$ &  &  & $\checkmark$ & $\checkmark$\\
		\hline
		\cite{bafghi2022degrees} 	& $\checkmark$ &  & $\checkmark$ &  & $\checkmark$ &  &  &  &   & $\checkmark$ & $\checkmark$ \\
		\hline
		\cite{Weihua}	& $\checkmark$ & $\checkmark$   & $\checkmark$ &  & $\checkmark$ &  & $\checkmark$ &  &  & $\checkmark$  & \\
		\hline
		\cite{Feng}	& $\checkmark$ &    &  &   &   & $\checkmark$ & $\checkmark$ & $\checkmark$ &  & $\checkmark$  & $\checkmark$ \\			\hline 
		\cite{chae2022cooperative}	& $\checkmark$ &   & $\checkmark$ &   &   &   &  $\checkmark$ & $\checkmark$  &  & $\checkmark$ & $\checkmark$\\
		\hline 
		\cite{zheng2023dof}	& $\checkmark$ &    &  &  & $\checkmark$ &  & $\checkmark$ &  &  & $\checkmark$ & \\
		\hline 
		\cite{OJCOMS}	& $\checkmark$ &    &  & $\checkmark$ &  &  &  &  &  & $\checkmark$ &\\
		\hline 
		\cite{nafea2021secure}	& $\checkmark$ &    &   &   &    &   & $\checkmark$ &  & $\checkmark$ & $\checkmark$ & $\checkmark$\\
		\hline 
		\cite{luo2024secure}	& $\checkmark$ &    &  &  &   & $\checkmark$  & $\checkmark$  &  & $\checkmark$  & $\checkmark$ & $\checkmark$\\
		\hline 
		\cite{Azari}	& $\checkmark$ &    & $\checkmark$ &  &  &  & $\checkmark$  &  & $\checkmark$ & $\checkmark$ & \\
		\hline 
		Ours	& $\checkmark$ &    &    &   &  $\checkmark$ &  & $\checkmark$ &  & $\checkmark$ & $\checkmark$ & $\checkmark$\\
		\hline 
	\end{tabular}
\end{table*}

$\text{Tx}_i$ intends to send a message $W_i$, which is chosen from $[1:2^{nR_{i}}]$, $i=1,2$, where $n$ represents the channel uses and $R_{i}$ is the transmission rate of the corresponding message.
    Based on received signal $\mathbf{y}_i$, $\text{Rx}_i$ can produce an estimate $\widehat{W}_i$ of the message $W_i$, $i=1,2$. A secrecy rate tuple $(R_{1},R_{2})$ is achievable if there exists a sequence of codes $(2^{nR_{1}},2^{nR_{2}},n)$ such that $
    \mathrm{Prob}(\widehat{W}_i \neq W_i) \leq \epsilon, i=1, 2$,
    $I(W_i; \mathbf{y}_\mathrm{e}) \leq \epsilon, i=1,2$,
    where $I(\cdot\ ;\ \cdot)$ represents the mutual information and $\epsilon$ vanishes as $n\rightarrow \infty$. This is known as \textit{strong security} \cite{mahdavifar2011achieving}. The secrecy capacity $(C_\text{s1},C_{s2})$ is defined as the closure of all achievable secrecy rate tuples. Thus, the sum-SDoF is defined as
    \begin{align}
        \text{sum-SDoF} \triangleq \lim_{P \rightarrow \infty} \sup (C_{\text{s}1}(P)+C_{\text{s}2}(P))/\log P,
    \end{align}
    where $P$ denotes the transmit power at each transmitter. The achievable sum-SDoF is a lower bound of sum-SDoF.

\section{Main Results and Discussion}

\begin{theorem}[Linear and Numerical Lower Bound]
	A linear and numerical sum-SDoF lower bound of an active RIS-assisted MIMO wiretap IC, defined in Section-II, is given by the solution of Problem (P0), shown on the top of next page.
	\begin{figure*}
		\begin{subequations}
		\begin{eqnarray}
		\text{(P0)}	 	\max_{\substack{
					f_{e1}, 
					f_{e2}, \\ f_{12},   
					f_{21} \in \mathbb{Z}^+ 
			}}  && \min\{N_2 - D_{21} + M_1  -D_{e1},  	M_2 - D_{12} + N_1 -D_{e2},  \min\{N_1,M_1-D_{e1}\} +  \min\{N_2,M_2-D_{e2}\}\}  \nonumber \\
	 \text{s.t.}		&&    D_{e1} = \min\{N_e, M_1\} - f_{e1}, \\
			&&    D_{e2} = \min\{N_e, M_2\} - f_{e2}, \\
			&&    D_{12} = \min\{N_1, M_2\} - f_{12}, \\
			&&      D_{21} = \min\{N_2, M_1\} - f_{21}, \\
			&& f_{e1} \max\{N_e, M_1\} + f_{e2} \max\{N_e, M_2\} + f_{12}\max\{N_1,M_2\} + f_{21}\max\{N_2,M_1\} \le R.
		\end{eqnarray}		
	\end{subequations} 
	\hrule	
	\end{figure*}
\end{theorem}

\begin{IEEEproof}
Please refer to  Section-IV. 
\end{IEEEproof}

	\begin{table}[t]
	\caption{Tabular Closed-Form with Symmetry Antenna Configurations}
	\label{tab2}
	\centering
	\begin{tabular}{|l|l|}
		\hline
		\text{\textbf{Case}} & \text{\textbf{Sum-SDoF Lower Bound}$/2$} \\
		\hline 
		$N \le M \, \&\, N_e \le M$ & $  \min \{M -N_e + \lfloor \frac{R - 2(M-N_e)M}{4M} \rfloor, N\}$ \\
		\hline 
		$N \le M \, \&\, N_e > M$ & $  \min \{\lfloor \frac{R}{2(N_e+M)} \rfloor, N\}$ \\
		\hline
		$N > M \, \& \, N_e \le M$ & $  \min \{M - N_e + \lfloor \frac{R - 2N(2M - N_e -N)}{2(M + N)} \rfloor, N\}$ \\
		\hline
		$N > M \, \& \, N_e > M$ & $  \min \{\lfloor \frac{R - 2N(M-N)}{2(N+N_e)} \rfloor, N\}$\\
		\hline
	\end{tabular}
\end{table}

\begin{remark}[The Idea of Proposed Scheme]
  To achieve the proposed sum-SDoF lower bound, we propose a unified transmission scheme for arbitrary antenna configurations, detailed in Section-IV. The idea of this scheme is integrating RIS interference and leakage elimination, ZF transmit beamforming, and receiver ID techniques together. Specifically, we adopt RIS beamforming to reduce the rank of interference and leakage matrices, yielding a closed-form relationship for matrices rank and number of RIS elements. After that, ZF transmit beamforming and receiver ID are taken into use with judiciously chosen configurations. Finally, we establish a sum-SDoF maximization problem, namely Problem (P0), by linear integer programming.
\end{remark}

\begin{remark}[Solution Algorithm to Problem (P0)]
	Note that Problem (P0) is a linear integer programming problem, which is NP-hard and non-convex. Thus, the optimal solution to Problem (P0) can be given by the BnB approach, which relies on bounding technique reduces the complexity of exhaustive search. We therefore invoke GROUBI software for solving this linear integer programming problem, where BnB is used with optimal solutions returned.	
\end{remark}

\begin{corollary}[Symmetry Antenna Configurations]
	If we assume that symmetry antenna configuration $M_1=M_2=M,\,N_1=N_2=N$, we can re-write Problem (P0) as
	\begin{subequations}
		\begin{eqnarray}
			\!\!\!\!\!\!\!\!\!\!\!\!\! (\text{P0}_1) \, \max_{\substack{
					t, f_{e}, \\
					f \in \mathbb{Z}^+
			}} && \!\!\!\!\!\!\!\! t \nonumber \\
			\text{s.t.} &&  \!\!\!\!\!\!\!\! t \le N - D + M  -D_{e}, \\
			&& \!\!\!\!\!\!\!\! t \le M - D + N -D_{e}, \\
			&& \!\!\!\!\!\!\!\! t \le 2\min\{N,M-D_{e}\},   \\
			&& \!\!\!\!\!\!\!\! D_e = \min\{N_e,M\} -f_e, \\
			&& \!\!\!\!\!\!\!\! D =  \min\{N,M\} - f, \\ 
			&& \!\!\!\!\!\!\!\! f_e \max\{N_e,M\} + f\max\{N,M\} \le \frac{R}{2}.  
		\end{eqnarray}
	\end{subequations} 
By analysis, we can obtain tabular closed-form sum-SDoF from Problem $(\text{P0}_1)$ in Table 2. Please refer to Appendix B for details of the derivation. 
\end{corollary}

	\begin{algorithm}[t]
	\caption{Algorithm for Minimizing Nuclear Norm to Rank Minimization}
	\begin{algorithmic}[1]
		\State \textbf{Input:} $\overline{\mathbf{H}} = \mathbf{H} + \mathbf{G} \mathbf{\Phi} \mathbf{D}$
		\State \textbf{Output:}  \text{min\_rank}
		\State \textbf{Initialization:} $\text{min\_rank} = \infty$ 
		\For{$\textit{Iteration} = 0$ \textbf{to} $\textit{Iteration} =\textit{Iteration}_{\max}$}   
		\State Solve optimization problem $\min_\mathbf{\Phi} \|\mathbf{H}\|_*$ using the splitting conic solver in CVXPY
		\If{Solution is successful}
		\State Obtain optimal $\mathbf{\Phi}^*$ from $\min_\mathbf{\Phi} \|\mathbf{H}\|_*$
		\State Compute $\mathbf{M}_{\text{optimal}} = \mathbf{H} + \mathbf{G} \mathbf{\Phi}^* \mathbf{D}$
		\State Set elements of $\mathbf{M}_{\text{optimal}}$ with absolute value less than $10^{-3}$ to $0$
		\State Compute singular values of $\mathbf{M}_{\text{optimal}}$
		\State Reconstruct $\mathbf{M}_{\text{modified}}$  by setting singular values less than $10^{-3}$ to $0$
		\State Compute the rank of $\mathbf{M}_{\text{modified}}$ as $D$
		\If{$D < \text{min\_rank}$}
		\State Update $\text{min\_rank} = D$	 
		\EndIf
		\Else
		\State Output failure message
		\EndIf
		\EndFor
		\State \textbf{Output:}   $\text{min\_rank}$
	\end{algorithmic}
\end{algorithm}

\begin{corollary}[Without Any Eavesdroppers]
	Besides, we can assume that there does not exist an eavesdropper. In this case, $D_{\mathrm{e}1} = D_{\mathrm{e}2} = f_{\mathrm{e}1} = f_{\mathrm{e}2} = N_\mathrm{e} = 0$. For this reduction, Problem (P0) can be re-written as 
		\begin{subequations}
		\begin{eqnarray}
			\!\!\!\! (\text{P0}_2) \, \max_{\substack{
					t, f_{12}, \\
					f_{21} \in \mathbb{Z}^+
			}} && \!\!\!\! t \nonumber \\
			\text{s.t.} &&  \!\!\!\!\! t \le N_2- \min\{N_2,M_1\} + M_1 + f_{21}, \\
			&& \!\!\!\! t \le N_1 - \min\{N_1,M_2\} + M_2 + f_{12}, \\
			&& \!\!\!\! t \le  \min\{N_1,M_1\} + \min\{N_2,M_2\},   \\
			&& \!\!\!\! f_{12} \max\{N_1,M_2\} + \nonumber \\   
			&& \!\!\!\! \qquad \qquad \quad f_{21}\max\{N_2,M_1\}   \le R. 
		\end{eqnarray}
	\end{subequations} 
	The above linear integer programming problem unifies that in \cite{zheng2023dof}, and can be optimally solved by GROUBI software using  BnB with heuristics  \cite{gurobi}.
\end{corollary}

\begin{theorem}[Linear and Numerical Upper Bound]
%It is divided into two cases based on whether leakage signals can be completely eliminated by RIS.	
 	
A linear and numerical sum-SDoF  upper bound of an active RIS-assisted MIMO wiretap IC, defined in Section-II, is given as follows:
\begin{eqnarray}
 && \!\!\!\!\!\!\!\!\!\!\!\! d_1 + d_2 \le \nonumber \\
 && \!\!\!\!\!\!\!\!\!\!\!\!	\begin{cases}
		\min\{M_1 + M_2, N_1 + N_2, M_1 + N_2 - D_{21},  \\
		\quad \quad   M_2 + N_1 - D_{12}\}, \qquad \,\, R > (M_1 + M_2)N_e,  \\
		\min\{M_1 + M_2-D_e, N_1 + N_2,\max\{M_1,N_2\},  \\ \quad \quad \,\,  \max\{M_2,N_1\}\},  \qquad \quad  R \le (M_1 + M_2)N_e.
			\end{cases}
			\label{eq6}
		\end{eqnarray}
where the rank of matrix $\overline{\mathbf{H}}_{ji}$ by denoted by $D_{ji}$ with  $\mathbf{\Phi} \in \mathbb{C}^{\overline{R} \times \overline{R}}$ and $\overline{R}  = R - (M_1 + M_2)N_e$, the rank of  matrix $\overline{\mathbf{H}}_{e} \triangleq [\mathbf{H}_{e1},\mathbf{H}_{e2}] + \mathbf{G}\mathbf{\Phi}[\mathbf{D}_{1},\mathbf{D}_{2}]$ is denoted by $D_{e}$. The value of those matrices' rank can be obtained from Algorithm 1.

%When $R \le (M_1 + M_2)N_e$, a linear and numerical sum-SDoF upper bound of an active RIS-assisted MIMO wiretap IC, defined in Section-II, is given as follows: 
%\begin{eqnarray}
%&& 	\!\!\!\!\!\!\!\! d_1 + d_2  \leq \min\{M_1 + M_2-D_e, N_1 + N_2, \nonumber \\
%	&& 	\!\!\!\!\!\!\!\! \qquad \qquad\qquad\,\,\, \max\{M_1,N_2\},  \max\{M_2,N_1\}\},  
%\end{eqnarray}
 
\end{theorem}

\begin{IEEEproof}
Theorem 2 distinguishes two cases based on the RIS’s ability to fully suppress leakage signals. For $R > (M_1 + M_2)N_e$, demonstrating the complete elimination of leakage signals through RIS, the detailed proof is provided in Section V. For proving $R \le (M_1 + M_2)N_e$, the first term comes from \cite[Theorem 2]{nafea2021secure} by treating MIMO wiretap IC as a MIMO wiretap channel with $M_1 + M_2$ antenna at the transmitter. The remaining three terms come from $R = (M_1 + M_2)N_e$ with complete leakage elimination, where the sum-SDoF is equivalent to the sum-DoF of MIMI IC \cite[Theorem 2]{jafar2007degrees}.
\end{IEEEproof}

\begin{figure*}[t]
	\centering
	\begin{subfigure}{0.245\linewidth}
		\centering
		\includegraphics[width=\linewidth, trim=14 8 25 20, clip]{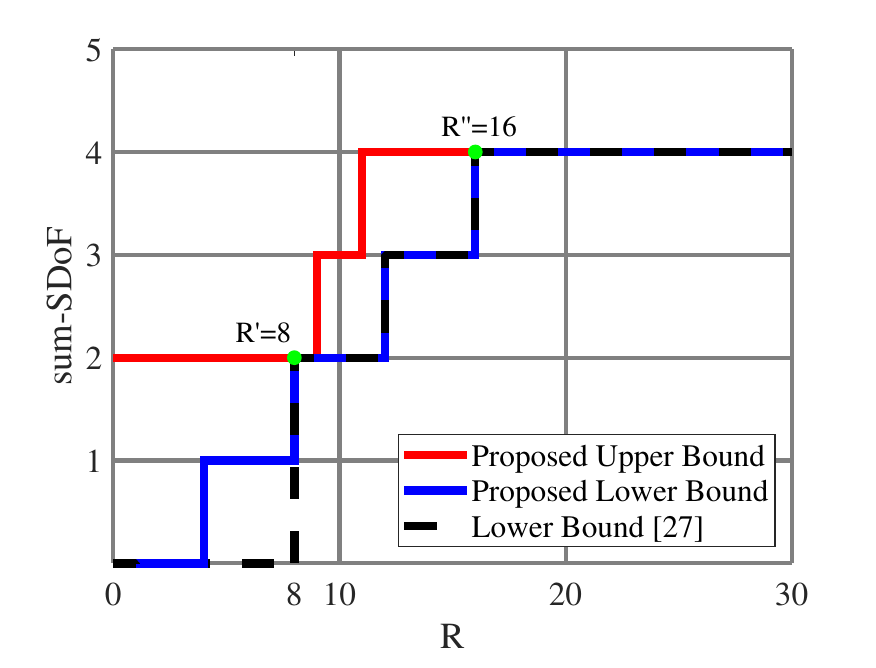}
		\caption{$(2,2,2,2,2)$}
		\label{fig:subfig1}
	\end{subfigure}%
	\begin{subfigure}{0.245\linewidth}
		\centering
		\includegraphics[width=\linewidth, trim=14 8 25 20, clip]{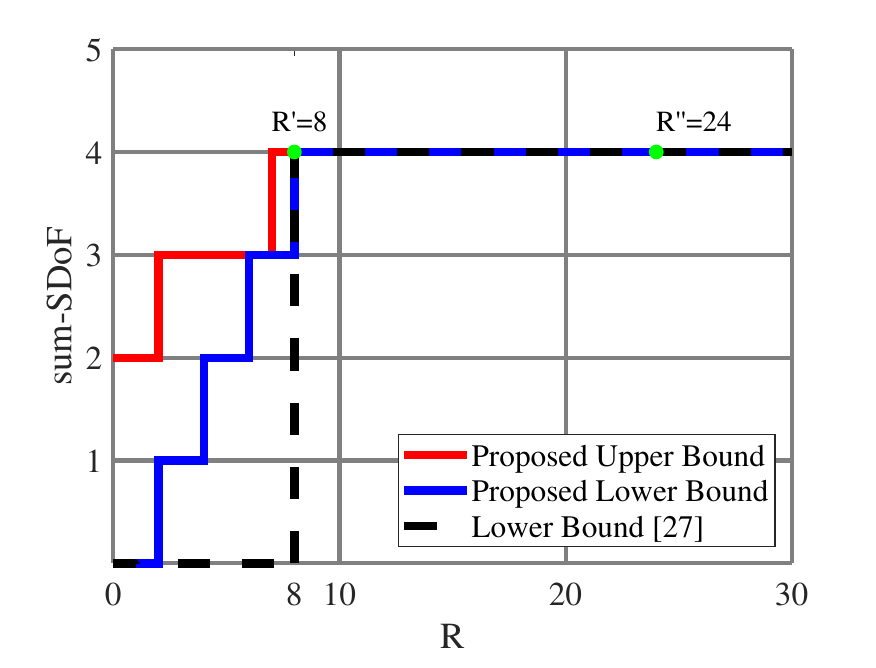}
		\caption{$(2,2,4,4,2)$}
		\label{fig:subfig2}
	\end{subfigure}%
	\begin{subfigure}{0.245\linewidth}
		\centering
		\includegraphics[width=\linewidth, trim=14 8 25 20, clip]{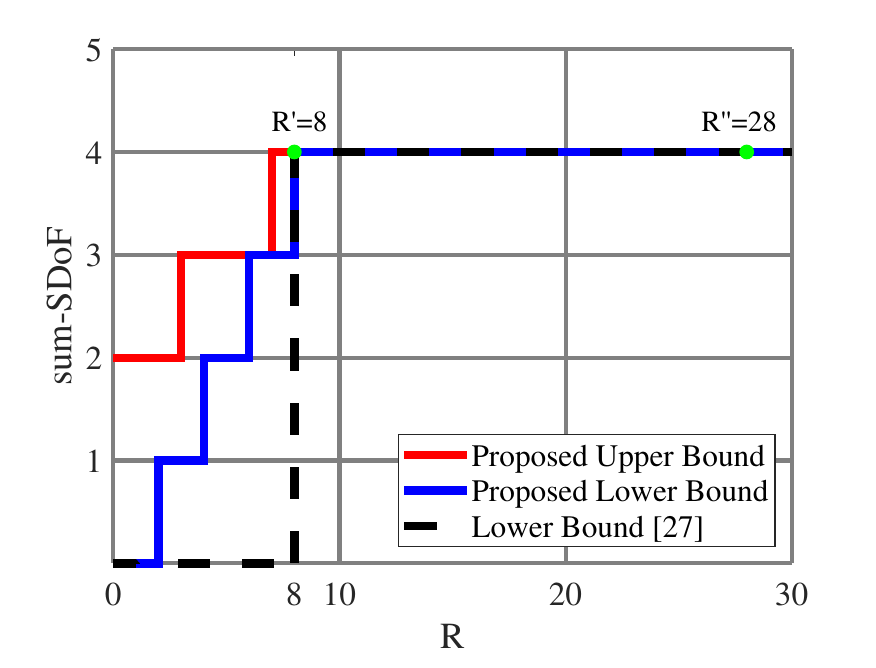}
		\caption{$(2,2,5,5,2)$}
		\label{fig:subfig3}
	\end{subfigure}%
	\begin{subfigure}{0.245\linewidth}
		\centering
		\includegraphics[width=\linewidth, trim=14 8 25 20, clip]{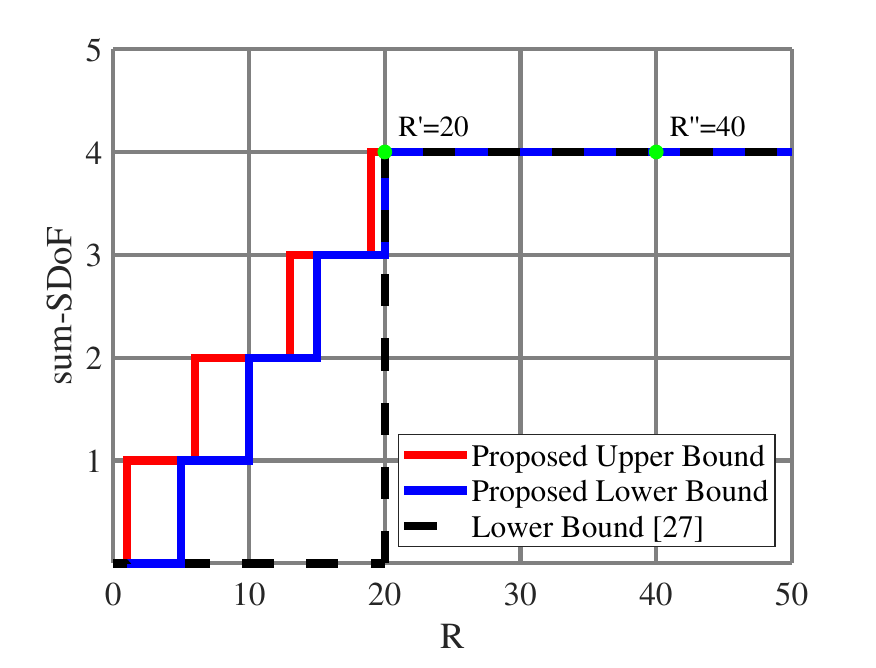}
		\caption{$(2,2,5,5,5)$}
		\label{fig:subfig4}
	\end{subfigure}
	
	\vspace{1ex}
	
	\begin{subfigure}{0.245\linewidth}
		\centering
		\includegraphics[width=\linewidth, trim=14 8 25 20, clip]{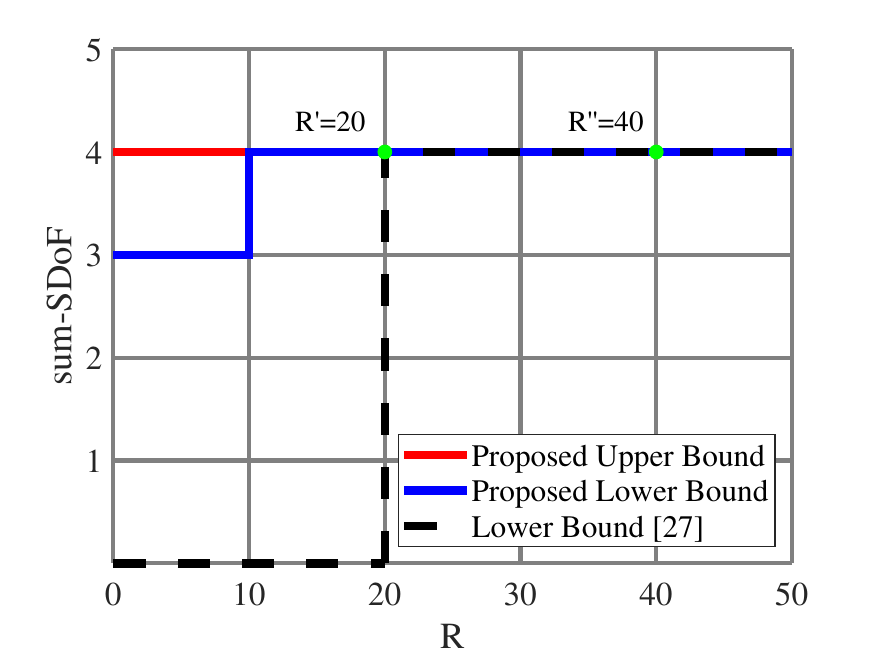}
		\caption{$(5,5,2,2,2)$}
		\label{fig:subfig5}
	\end{subfigure}%
	\begin{subfigure}{0.245\linewidth}
		\centering
		\includegraphics[width=\linewidth, trim=14 8 25 20, clip]{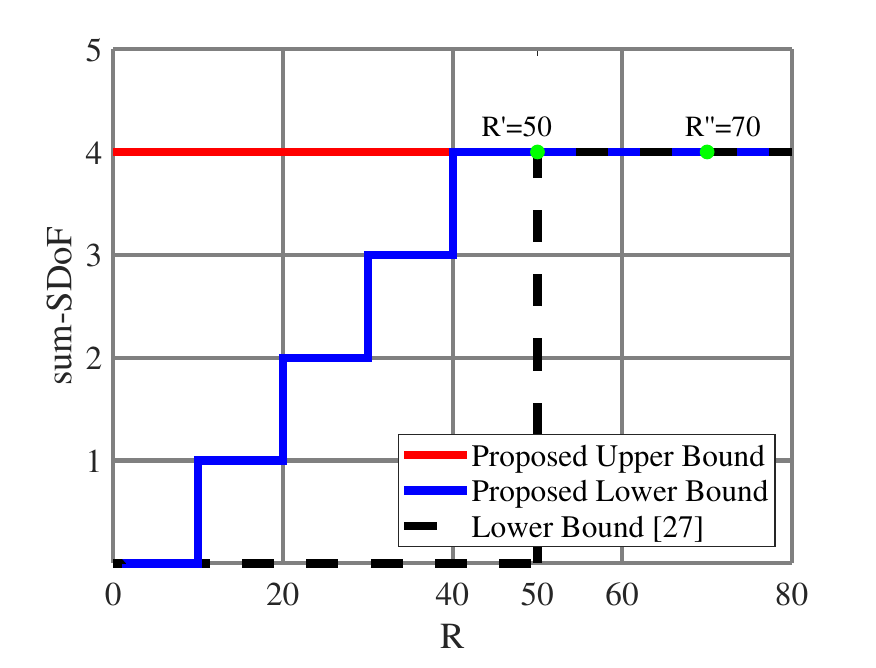}
		\caption{$(5,5,2,2,5)$}
		\label{fig:subfig6}
	\end{subfigure}
	\begin{subfigure}{0.245\linewidth}
		\centering
		\includegraphics[width=\linewidth, trim=14 8 25 20, clip]{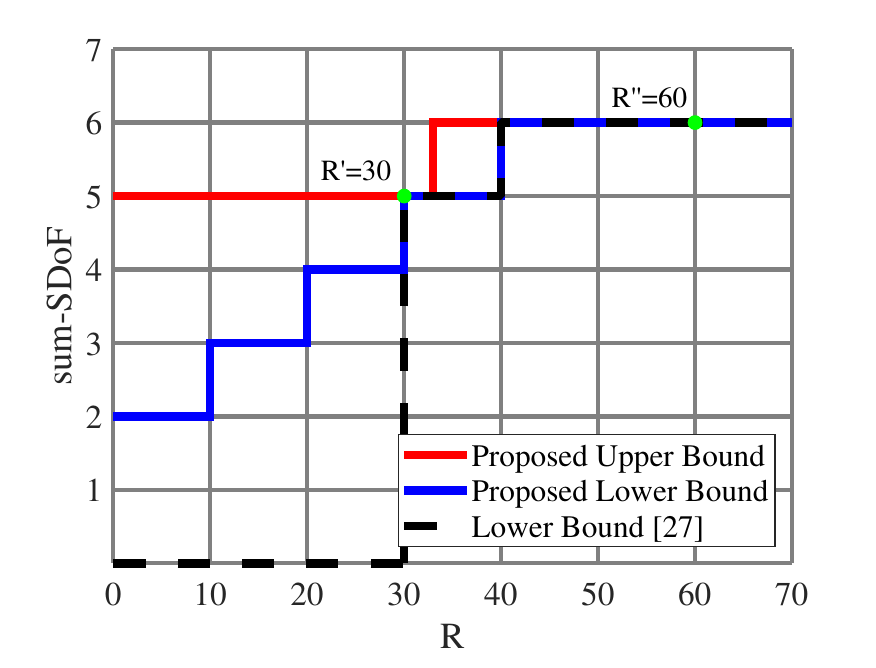}
		\caption{$(5,5,3,3,3)$}
		\label{fig:subfig7}
	\end{subfigure}%
	\begin{subfigure}{0.245\linewidth}
		\centering
		\includegraphics[width=\linewidth, trim=14 8 25 20, clip]{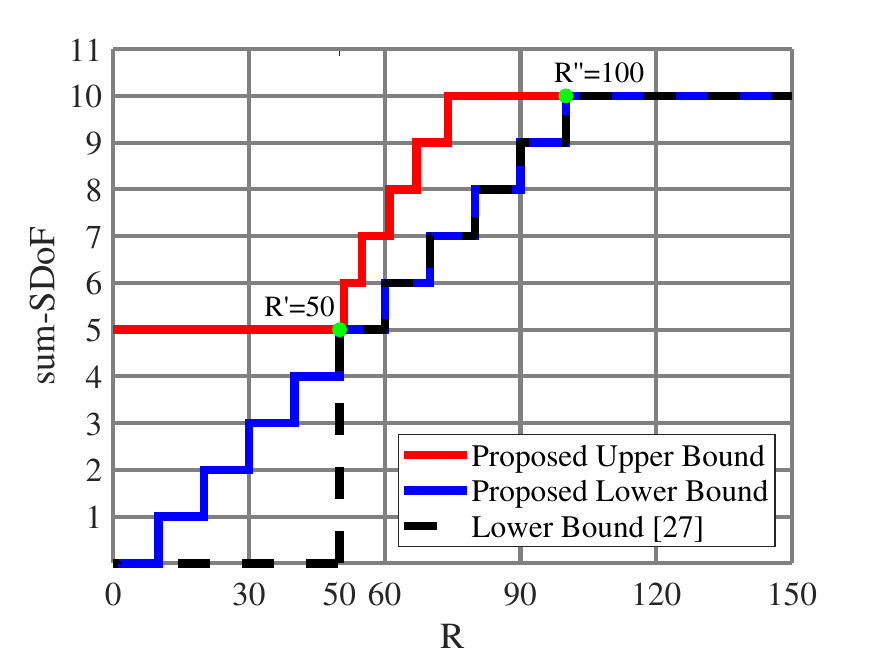}
		\caption{$(5,5,5,5,5)$}
		\label{fig:subfig8}
	\end{subfigure}
	
		\centering
	\begin{subfigure}{0.245\linewidth}
		\centering
		\includegraphics[width=\linewidth, trim=14 8 25 20, clip]{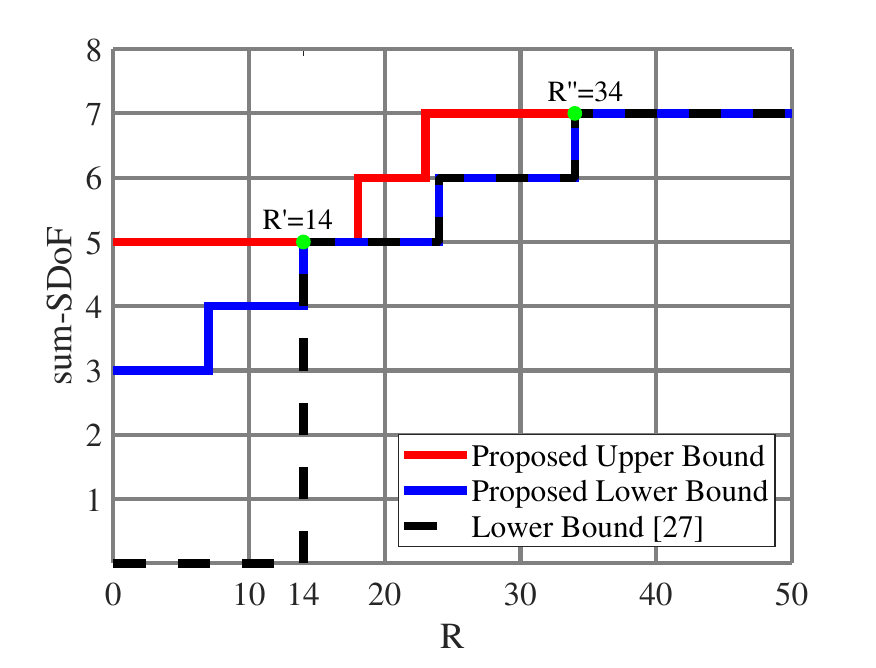}
		\caption{$(5,2,5,2,2)$}
		\label{fig:subfig9}
	\end{subfigure}%
	\begin{subfigure}{0.245\linewidth}
		\centering
		\includegraphics[width=\linewidth, trim=14 8 25 20, clip]{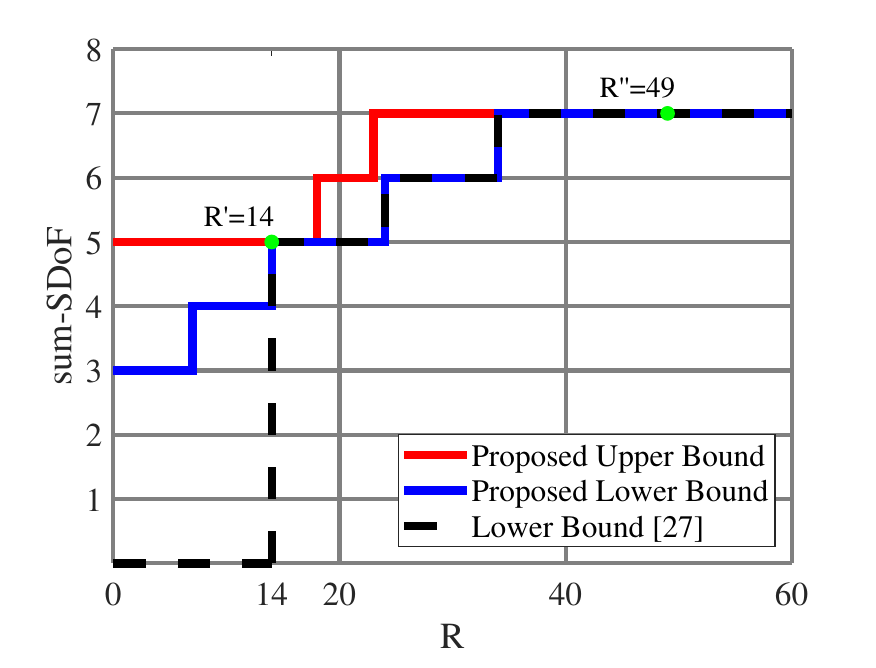}
		\caption{$(5,2,5,5,2)$}
		\label{fig:subfig10}
	\end{subfigure}%
	\begin{subfigure}{0.245\linewidth}
		\centering
		\includegraphics[width=\linewidth, trim=14 8 25 20, clip]{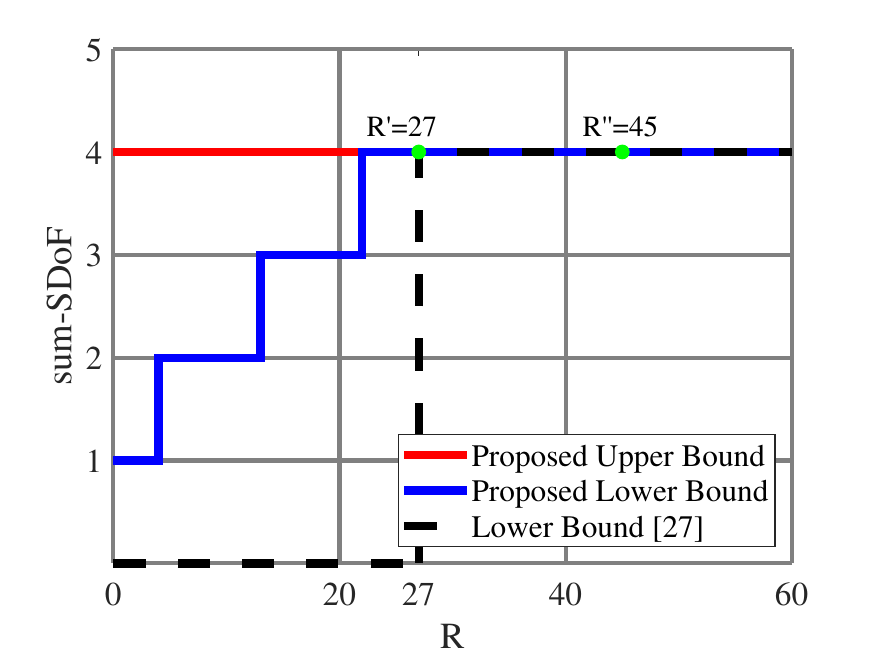}
		\caption{$(5,4,2,2,3)$}
		\label{fig:subfig11}
	\end{subfigure}%
	\begin{subfigure}{0.245\linewidth}
		\centering
		\includegraphics[width=\linewidth, trim=14 8 25 20, clip]{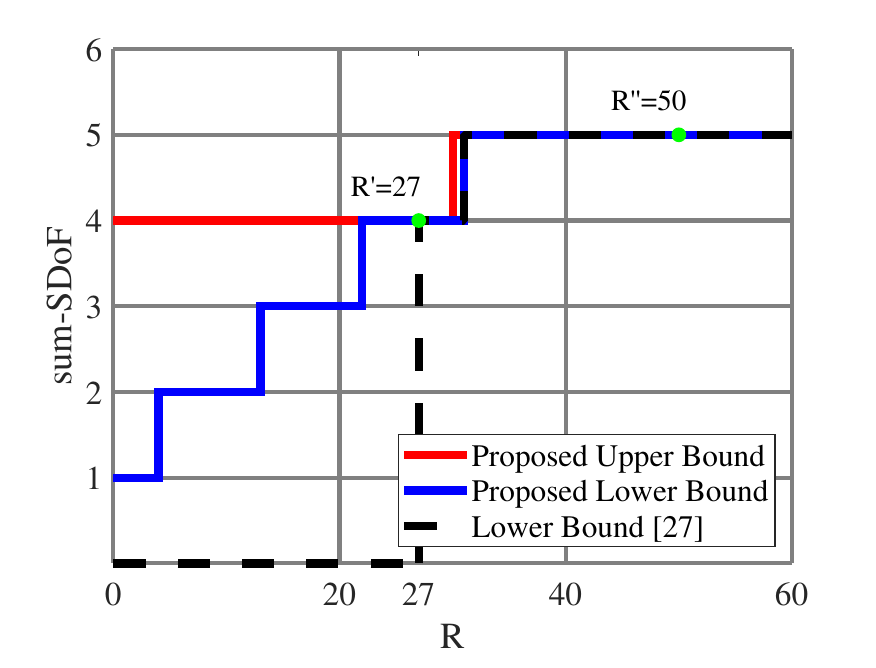}
		\caption{$(5,4,2,3,3)$}
		\label{fig:subfig12}
	\end{subfigure}
	
	\vspace{1ex}
	
	\begin{subfigure}{0.245\linewidth}
		\centering
		\includegraphics[width=\linewidth, trim=14 8 25 20, clip]{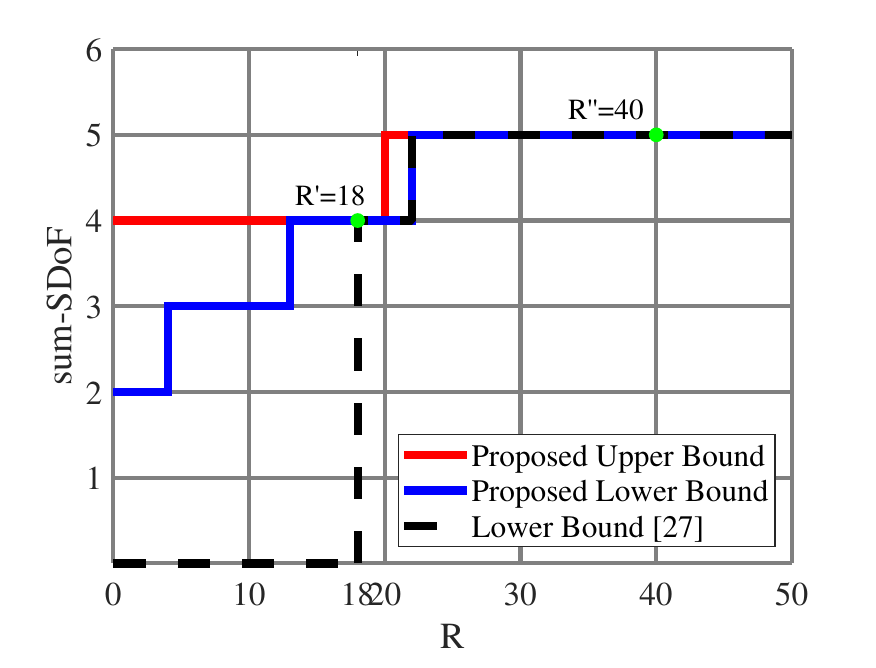}
		\caption{$(5,4,3,2,2)$}
		\label{fig:subfig13}
	\end{subfigure}%
	\begin{subfigure}{0.245\linewidth}
		\centering
		\includegraphics[width=\linewidth, trim=14 8 25 20, clip]{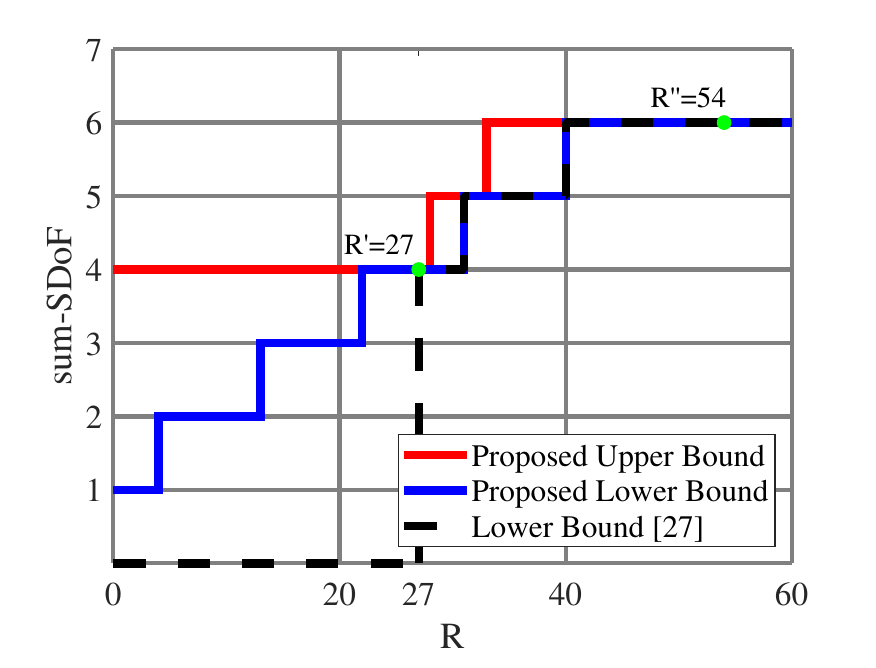}
		\caption{$(5,4,3,3,3)$}
		\label{fig:subfig14}
	\end{subfigure}
	\begin{subfigure}{0.245\linewidth}
		\centering
		\includegraphics[width=\linewidth, trim=14 8 25 20, clip]{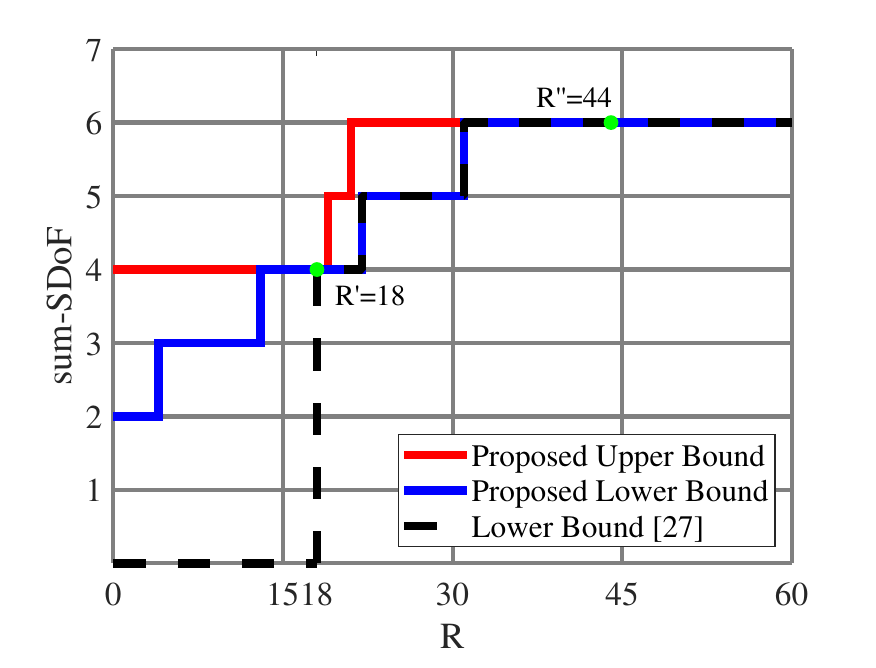}
		\caption{$(5,4,4,2,2)$}
		\label{fig:subfig15}
	\end{subfigure}%
	\begin{subfigure}{0.245\linewidth}
		\centering
		\includegraphics[width=\linewidth, trim=14 8 25 20, clip]{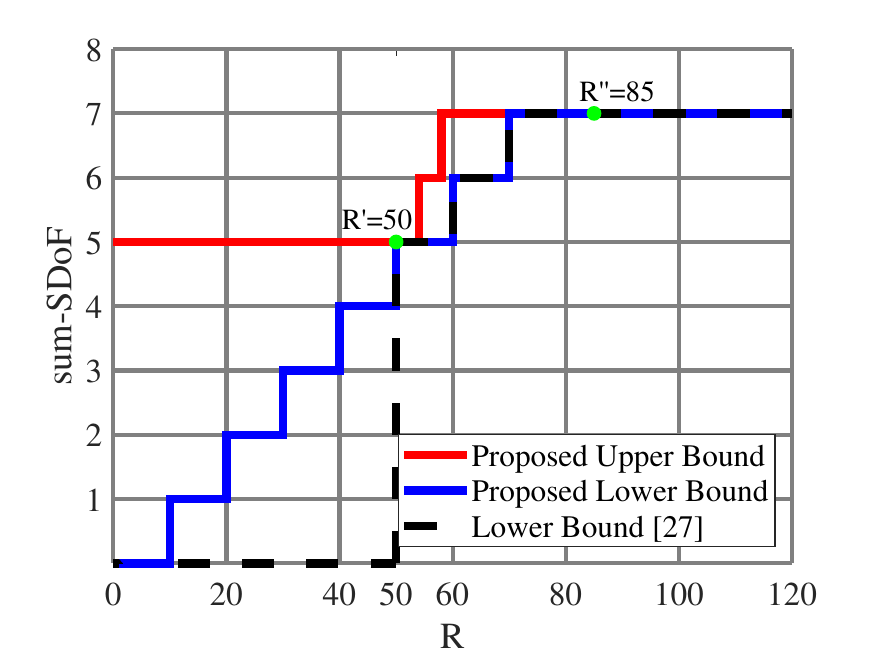}
		\caption{$(5,5,2,5,5)$}
		\label{fig:subfig16}
	\end{subfigure}
	\caption{Comparison of lower and upper bounds, where the antenna configuration is denoted by $(M_1,M_2,N_1,N_2,N_e)$.} \label{F2}
	\label{fig:main}
\end{figure*}
 
\begin{remark}[Performance Evaluation of Bounds]
	We compare our proposed lower and upper bounds, and the lower bound in \cite{zheng2023dof} numerically. Fig. \ref{F2} shows the following:
	
 The existing lower bound in \cite{zheng2023dof} aligns with our proposed lower bound with $R > (M_1 + M_2)N_e$ since leakage signals are completely canceled. When $R \leq (M_1 + M_2)N_e$ (all leakage links elimination), denoted as $R'$ in the figure, the lower bound in \cite{zheng2023dof} remains zero due to inevitable information leakage. By contrast, our proposed lower bound is non-zero and higher in this regime, which demonstrates the effectiveness and superiority of the proposed design. Additionally, it is observed that the two lower bounds are guaranteed to reach the upper bound with $R\ge (M_1 + M_2)N_e+M_1N_2+M_2N_1$ (all leakage and interference links elimination), denoted as $R''$ in the figure. In this case, all leakage and interference is canceled by RIS beamforming thus the MIMO wiretap IC degenerates to two point-to-point MIMO channels and the exact sum-SDoF can be achieved.  
 
  Nevertheless, it's observed that our proposed bounds might align with each other when $R< (M_1 + M_2)N_e+M_1N_2+M_2N_1$. The results in Figs. \ref{fig:subfig2}, \ref{fig:subfig3}, and \ref{fig:subfig4} indicate that proposed lower and upper bounds coincides with each other when $R > (M_1 + M_2)N_e$. In this case, the information leakage is avoided by utilizing RIS beamforming and interference can be fully eliminated via ZF or ID. For results shown in Figs. \ref{fig:subfig1}, \ref{fig:subfig8}, and \ref{fig:subfig9} with symmetry antenna configurations $M_1=N_1,M_2=N_2$, it is observed that the minimal $R$ to achieve the upper bound must be larger than $(M_1 + M_2)N_e+M_1N_2+M_2N_1$ because the given antenna configurations do not provide any ZF or ID capability. The results in Figs. \ref{fig:subfig7}, \ref{fig:subfig10}, and \ref{fig:subfig12}-\ref{fig:subfig16} show that, the minimal $R$ required to achieve the upper bound is between $(M_1 + M_2)N_e$ and $(M_1 + M_2)N_e+M_1N_2+M_2N_1$ for asymmetry antenna configurations, which reveals that RIS beamforming is necessary to assist ZF and ID to eliminate all interference. 
   
	Overall, our lower bound meets or exceeds the existing bound in \cite{zheng2023dof}, since our scheme ensures secure transmission, in contrast to \cite{zheng2023dof}, where transmission is suspended if security cannot be guaranteed. In addition, our upper bound coincides with our lower bound  for many antenna configurations, including RIS configurations. This validates the effectiveness of the proposed scheme, associated optimization Problem (P0), and nuclear norm minimization.
\end{remark}

\begin{remark}[Discussion on Algorithm 1]
	Algorithm 1 aims to minimize the nuclear norm, namely the sum of singular values, for rank minimization. Fortunately, our problem is convex,  because the nuclear norm is convex and 
	$\mathbf{H} + \mathbf{G} \mathbf{\Phi} \mathbf{D}$ is affine, meaning their composition preserves convexity. However, the solution of splitting conic solver in CVXPY may not be equal to the exact minimum rank of $\overline{\mathbf{H}}$. This is because nuclear norm is a convex envelope of rank \cite{Papailiopoulos}. To overcome this limitation, we repeatedly invoke the splitting conic solver as many as $\textit{Iteration}_{\max}$ times, and then select best results among them.
\end{remark}

%\mathbf{Corollary 1.}

\section{Proof of Theorem 1}

\subsection{The Proposed Scheme and Associated Sum-SDoF Lower Bound Maximization Problem}

In our proposed scheme, there are three steps regarding to the scheme design.

\textit{Step-I (SVD Transformation):} We first use  brief notations on the cascaded channel. To this end, we denote leakage channels as $\overline{\mathbf{H}}_{ei} \triangleq \mathbf{H}_{\mathrm{e}i}+\mathbf{G}_{\mathrm{e}}\mathbf{\Phi }\mathbf{D}_{i},\,i=1,2$, and interference and desired signal channels as $\overline{\mathbf{H}}_{ji} \triangleq \mathbf{H}_{ji}+\mathbf{G}_{j}\mathbf{\Phi }\mathbf{D}_{i},\,i,j=1,2$. Using these short notations, we have the following input-output relationship, 
\begin{subequations}
	\begin{eqnarray}
	&&	\mathbf{y}_\mathrm{e} = \overline{\mathbf{H}}_{\mathrm{e}1}\mathbf{P}_1\mathbf{x}_1 + 
		\overline{\mathbf{H}}_{\mathrm{e}2}\mathbf{P}_2\mathbf{x}_2, \\
	&&	\mathbf{y}_1 = \overline{\mathbf{H}}_{11}\mathbf{P}_1\mathbf{x}_1 + 
		\overline{\mathbf{H}}_{12}\mathbf{P}_2\mathbf{x}_2, \\
	&&	\mathbf{y}_2 = \overline{\mathbf{H}}_{21}\mathbf{P}_1\mathbf{x}_1 + 
		\overline{\mathbf{H}}_{22}\mathbf{P}_2\mathbf{x}_2,
	\end{eqnarray}
\end{subequations}
where $\overline{\mathbf{H}}_{\mathrm{e}i} \in \mathbb{C}^{N_e \times M_i},\,i=1,2$,  $\overline{\mathbf{H}}_{ji} \in \mathbb{C}^{N_j \times M_i},\,i,j=1,2$, $\mathbf{P}_i \in \mathbb{C}^{M_i \times \delta_i},\,i=1,2$, and $\mathbf{x}_i \in \mathbb{C}^{\delta_i \times 1},\,i=1,2$. Note that $\delta_i$ denotes the amount of data symbols transmitted at $\text{Tx}_i$, which will be designated later on. Next, we perform SVD of the matrices of IC, leading to the following input-output relationship, 
\begin{subequations}
	\begin{eqnarray}
		&&	\mathbf{y}_\mathrm{e} = \overline{\mathbf{H}}_{\mathrm{e}1}\mathbf{P}_1\mathbf{x}_1 + 
		\overline{\mathbf{H}}_{\mathrm{e}2}\mathbf{P}_2\mathbf{x}_2, \label{4a}\\
		&&	\mathbf{y}_1 = \overline{\mathbf{H}}_{11}\mathbf{P}_1\mathbf{x}_1 + 
	\mathbf{U}_{12}{\bf{\Sigma}}_{12}\mathbf{V}^T_{12}\mathbf{P}_2\mathbf{x}_2, \label{4b} \\
		&&	\mathbf{y}_2 = \mathbf{U}_{21}{\bf{\Sigma}}_{21}\mathbf{V}^T_{21}\mathbf{P}_1\mathbf{x}_1 + 
		\overline{\mathbf{H}}_{22}\mathbf{P}_2\mathbf{x}_2, \label{4c}
	\end{eqnarray}
\end{subequations}
where unitary matrices $\mathbf{U}_{12} \in \mathbb{C}^{N_1 \times N_1}$, $\mathbf{V}_{12} \in \mathbb{C}^{M_2 \times M_2}$, $\mathbf{U}_{21} \in \mathbb{C}^{N_2 \times N_2}$, $\mathbf{V}_{21} \in \mathbb{C}^{M_1 \times M_1}$, and diagonal matrices for singular values ${\bf{\Sigma}}_{12} \in \mathbb{C}^{N_1 \times M_2}$ and  ${\bf{\Sigma}}_{21} \in \mathbb{C}^{N_2 \times M_1}$. Note that ${\bf{\Sigma}}_{12}$ and  ${\bf{\Sigma}}_{21}$ rank are with $D_{12}$ and $D_{21}$, respectively. Through $\mathbf{y}_1' = \mathbf{U}_{12}^H\mathbf{y}_1, \overline{\mathbf{H}}_{11}' = \mathbf{U}_{12}^H\overline{\mathbf{H}}_{11},$ $\mathbf{y}_2' = \mathbf{U}_{21}^H\mathbf{y}_2, \overline{\mathbf{H}}_{22}' = \mathbf{U}_{21}^H\overline{\mathbf{H}}_{22},$ we can transform \eqref{4a}-\eqref{4c} into 
\begin{subequations}
	\begin{eqnarray}
		&&	\mathbf{y}_\mathrm{e} = \overline{\mathbf{H}}_{\mathrm{e}1}\mathbf{P}_1\mathbf{x}_1 + 
		\overline{\mathbf{H}}_{\mathrm{e}2}\mathbf{P}_2\mathbf{x}_2, \\
		&&	\mathbf{y}_1' = \overline{\mathbf{H}}_{11}'\mathbf{P}_1\mathbf{x}_1 + 
	 {\bf{\Sigma}}_{12}\mathbf{V}^T_{12}\mathbf{P}_2\mathbf{x}_2, \\
		&&	\mathbf{y}_2' =  {\bf{\Sigma}}_{21}\mathbf{V}^T_{21}\mathbf{P}_1\mathbf{x}_1 + 
		\overline{\mathbf{H}}_{22}'\mathbf{P}_2\mathbf{x}_2.
	\end{eqnarray}
\end{subequations}
To maximize the sum-SDoF achieved and ensure zero leakage, we need to design the transmit beamforming matrices $\mathbf{P}_i,\,i=1,2$ and RIS beamforming matrix ${\bf{\Phi}}$, which will be addressed in Step-II. 

\textit{Step-II (Design of Beamforming Matrices):} We then consider 
the transmit beamforming matrices and RIS beamforming matrix design. In order to enlarge the capabilities of ZF and ID following RIS beamforming, the design principle of RIS beamforming is to reduce the rank of leakage and interference matrices by explicitly eliminating their rows/columns.
To this end, we establish the following linear equations, 
\begin{subequations}
	\begin{eqnarray}
&& \!\!\!\!\!\!\!\!\!\!\!\!\!\!  	\text{vec}(\underbrace{\mathbf{H}_{\mathrm{e}i}+\mathbf{G}_{\mathrm{e}}\mathbf{\Phi }\mathbf{D}_{i}}_{\overline{\mathbf{H}}_{ei}}) = \begin{bmatrix}
			\mathbf{0}_{ f_{ei}\max\{N_e,M_i\}}\\ 
			\mathbf{*}
		\end{bmatrix},  
	i=1,2, \label{6a} \\
&&	\!\!\!\!\!\!\!\!\!\!\!\!\!\!  	\text{vec}(\underbrace{\mathbf{H}_{ji}+\mathbf{G}_{j}\mathbf{\Phi }\mathbf{D}_{i}}_{\overline{\mathbf{H}}_{ji}}) = \begin{bmatrix}
		\mathbf{0}_{ f_{ji}\max\{N_j,M_i\}}\\ 
		\mathbf{*}
	\end{bmatrix}, \nonumber \\ 
&&\!\!\!\!\!\!\!\!\!\!\!\!\!\!  \qquad \qquad \qquad \qquad \qquad \qquad \qquad  	i,j=1,2,\,i\ne j, \label{6b}
	\end{eqnarray}
\end{subequations}
where $*$ denotes non-zero elements. Obviously, \eqref{6a} stands for 
the elimination of $f_{ei}$ rows if $N_e \leq M_i$ or the elimination of $f_{ei}$ columns if $M_i < N_e$.  Likewise, \eqref{6b} stands for the elimination of $f_{ji}$ rows if $N_j \leq M_i$ or the elimination of $f_{ji}$ columns if $M_i < N_j$.
This leads to the following rank reduction relationship, 
\begin{subequations}
	\begin{eqnarray}
&&		D_{e1} = \min\{N_e, M_1\} - f_{e1},  \label{7a} \\
&&      D_{e2} = \min\{N_e, M_2\} - f_{e2}, \label{7b} \\
&&		D_{12} = \min\{N_1, M_2\} - f_{12}, \label{7c}\\
&&		D_{21} = \min\{N_2, M_1\} - f_{21}, \label{7d}
	\end{eqnarray}
\end{subequations}
where $D_{e1}$, $D_{e2}$, $D_{12}$, $D_{21}$ denote the rank of matrices $\overline{\mathbf{H}}_{\mathrm{e}1}$, $\overline{\mathbf{H}}_{\mathrm{e}2}$, $\overline{\mathbf{H}}_{12}$, $\overline{\mathbf{H}}_{21}$, respectively. In order to solve \eqref{6a}-\eqref{6b}, due to Appendix A, the following relationship should satisfy
\begin{eqnarray}
&&	\!\!\!\!\!\!\!\! f_{e1} \max\{N_e, M_1\} + f_{e2} \max\{N_e, M_2\} +  \nonumber  \\ && \quad \!\!\!\!\!\!\!\! f_{12}\max\{N_1,M_2\} + f_{21}\max\{N_2,M_1\} \le R.  \label{8}
\end{eqnarray}

Under these rank-deficient leakage and interference matrices, we then design the transmit beamforming matrices with reference to \cite{Krishnamurthy}. Foremost, we need to null all leakage signals out, i.e.,
\begin{equation}
	\overline{\mathbf{H}}_{\mathrm{e}1}\mathbf{P}_1 = \mathbf{0}, \quad 	\overline{\mathbf{H}}_{\mathrm{e}2}\mathbf{P}_2 = \mathbf{0}. 
\end{equation} 
Equivalently, $\mathbf{P}_1 \in \mathbb{C}^{M_1 \times \delta_1}$ and $\mathbf{P}_2 \in \mathbb{C}^{M_2 \times \delta_2}$ must locate at the null-space of $\overline{\mathbf{H}}_{\mathrm{e}1}$ and $\overline{\mathbf{H}}_{\mathrm{e}2}$, respectively.   
Then, there can be two attempts regarding to the order of ZF and ID used for interference signals. 

\textit{Attempt 1:} Let $\delta_1 = \min\{N_1,M_1 - D_{e1}\}$, i.e., $\text{Tx}_1$ transmits at its maximal ability. Regarding this, we design the transmission at $\text{Tx}_2$ by using ZF first then ID. To be specific, let the first $M_1 - \delta_1 -D_{e1}$ elements in $\mathbf{V}^T_{21}\mathbf{P}_1$ be zero, which can be done by solving
\begin{equation}
\begin{bmatrix}
	\mathbf{V}^T_{21}\\
	\overline{\mathbf{H}}_{\mathrm{e}1}
\end{bmatrix}\mathbf{P}_1 	 = \begin{bmatrix}
		\mathbf{0}_{M_1 - \delta_1},	\mathbf{*},  
		\mathbf{0}_{N_e}
	\end{bmatrix}^T.  
\end{equation}
After ZF, it can be seen that there are $D_{21} - (M_1 - \delta_1 -D_{e1})$ interference at $\text{Tx}_2$. The above interference is then dealt with ID at  $\text{Tx}_2$. As such, $\text{Tx}_2$ should transmit $N_2 - (D_{21} - (M_1 - \delta_1 -D_{e1}))$ symbols, namely $\delta_2 = N_2 - (D_{21} - (M_1 - \delta_1 -D_{e1}))$. 

\textit{Attempt 2:} Let $\delta_1 = \min\{N_1,M_1 - D_{e1}\}$, i.e., $\text{Tx}_1$ transmits at its maximal ability. Regarding this, we design the transmission at $\text{Tx}_2$ by using ID first then ZF. To be specific, let the first $N_1 - \delta_1$ elements in $\mathbf{V}^T_{12}\mathbf{P}_2$ be non-zero, which is dealt with ID. Then, we suffice to null  remaining $D_{12} - (N_1 - \delta_1)$ interference at $\text{Tx}_1$ out. This can be done by solving
\begin{equation}
	\begin{bmatrix}
		\mathbf{V}^T_{12}\\
		\overline{\mathbf{H}}_{\mathrm{e}2}
	\end{bmatrix}\mathbf{P}_2 	 = \begin{bmatrix}
		\mathbf{*}, 	\mathbf{0}_{D_{12} - (N_1 - \delta_1)}, 
		\mathbf{0}_{N_e}
	\end{bmatrix}^T.  
\end{equation}
As such, $\text{Tx}_2$ should transmit $M_2 - (D_{12} - (N_1 - \delta_1)) -D_{e2}$ symbols, namely $\delta_2 = M_2 - (D_{12} - (N_1 - \delta_1))-D_{e2}$.

Furthermore, the number of symbols transmitted at $\text{Tx}_2$ should satisfy the maximum transmission and decoding ability, namely $\min\{N_2,M_2-D_{e2}\}$. Therefore, we have $\delta_2 \le \min\{N_2,M_2-D_{e2}\}$. To obtain the exact sum-SDoF achieved, we next formulate and solve a sum-SDoF lower bound maximization problem.

\textit{Step-III (Optimization):} Because from Step-II, we have
\begin{equation}
	\delta_1 + \delta_2 \le   
	\begin{cases}
		N_2 - D_{21} + M_1  -D_{e1}, \\ 
		M_2 - D_{12} + N_1 -D_{e2}, \\
		\min\{N_1,M_1-D_{e1}\} +    \min\{N_2,M_2-D_{e2}\},
	\end{cases} \nonumber 
\end{equation}
the sum-SDoF lower bound maximization problem can be formulated as follows,
\begin{eqnarray}
\!\!\!\!\!\!\!\!\!\!\!\! (\text{P0$'$}) \,\,\,	\max_{\substack{
		  f_{e1}, 
			f_{e2},  f_{12},   \\
			f_{21} \in \mathbb{Z}^+, t
	}} && \!\!\!\!\!\! t \nonumber \\
	\text{s.t.} &&  \!\!\!\!\!\! t \le N_2 - D_{21} + M_1  -D_{e1}, \\
	&& \!\!\!\!\!\! t \le M_2 - D_{12} + N_1 -D_{e2}, \\
	&& \!\!\!\!\!\! t \le \min\{N_1,M_1-D_{e1}\} + \nonumber \\
	&& \!\!\!\!\!\! \qquad \qquad \min\{N_2,M_2-D_{e2}\}, \\
	&& \!\!\!\!\!\! \eqref{7a}-\eqref{7d},\,\eqref{8}.
\end{eqnarray}
It can be seen that Problem (P0$'$) is equivalent to Problem (P0), thus Theorem 1 is proven.

\section{Proof of Theorem 2}

In order to prove $R > (M_1 + M_2)N_e$, leakage signals can be completely canceled by RIS. Thus, $\overline{R} = R - (M_1 + M_2)N_e$ RIS elements are used for eliminating interference, which is equivalent to that in MIMO IC. In this case, we therefore first utilize the result of in \cite[Theorem 4]{Tang}. This is because, at $R > (M_1 + M_2)N_e$, all leakage signals can be removed by the RIS and it is equivalent to a MIMO IC without eavesdropping. Then, by means of in \cite[Theorem 4]{Tang}, we discuss case by case for deriving the desired upper bound. The details of this proof are illustrated as follows:

According to \cite[Theorem 4]{Tang}, we have
\begin{align}
	d_1 + d_2 &\leq \text{rk} \left( \begin{bmatrix} \mathbf{H}_{11} & \mathbf{H}_{12} \end{bmatrix} \right) + \text{rk} \left( \begin{bmatrix} \mathbf{H}_{12} \\ \mathbf{H}_{22} \end{bmatrix} \right) - \text{rk}(\mathbf{H}_{12}), \ \label{EE01} \\
	d_1 + d_2 &\leq \text{rk} \left( \begin{bmatrix} \mathbf{H}_{22} & \mathbf{H}_{21} \end{bmatrix} \right) + \text{rk} \left( \begin{bmatrix} \mathbf{H}_{11} \\ \mathbf{H}_{21} \end{bmatrix} \right) - \text{rk}(\mathbf{H}_{21}) . \label{EE02}
\end{align}

For the first part, we divide \eqref{EE01} into two cases, and then expand them into the following.

For $N_1 \leq M_2$ Case, we have
\begin{subequations}
	\begin{eqnarray}
		&& \!\!\!\!\!\!\!\!\! \text{rk} \left( \begin{bmatrix} \mathbf{H}_{11} & \mathbf{H}_{12} \end{bmatrix} \right) = \min \{N_1,M_1+M_2\} = N_1, \\
		&& \!\!\!\!\!\!\!\!\! \text{rk} \left( \begin{bmatrix} \mathbf{H}_{12} \\ \mathbf{H}_{22} \end{bmatrix} \right) = \min \{M_2,D_{12}+N_2\}, \\
		&& \!\!\!\!\!\!\!\!\! \text{rk} \left( \begin{bmatrix} \mathbf{H}_{11} & \mathbf{H}_{12} \end{bmatrix} \right) + \text{rk} \left( \begin{bmatrix} \mathbf{H}_{12} \\ \mathbf{H}_{22} \end{bmatrix} \right) - \text{rk}(\mathbf{H}_{12}) \nonumber \\
		&&\!\!\!\!\!\!\!\!\! \quad = \min \{N_1+M_2-D_{12},N_1+N_2\}.\label{EE03}
	\end{eqnarray}
\end{subequations}

Otherwise, for $N_1 > M_2$ Case, we have
\begin{subequations}
	\begin{eqnarray}
		&&\!\!\!\!\!\!\!\!\!\!\!\!\!\!\!\!\!\!\text{rk} \left( \begin{bmatrix} \mathbf{H}_{11} & \mathbf{H}_{12} \end{bmatrix} \right) = \min \{N_1,M_1+D_{12}\}, \\
		&&\!\!\!\!\!\!\!\!\!\!\!\!\!\!\!\!\!\!\text{rk} \left( \begin{bmatrix} \mathbf{H}_{12} \\ \mathbf{H}_{22} \end{bmatrix} \right) = \min \{M_2,N_1+N_2\} = M_2, \\
		&&\!\!\!\!\!\!\!\!\!\!\!\!\!\!\!\!\!\!\text{rk} \left( \begin{bmatrix} \mathbf{H}_{11} & \mathbf{H}_{12} \end{bmatrix} \right) + \text{rk} \left( \begin{bmatrix} \mathbf{H}_{12} \\ \mathbf{H}_{22} \end{bmatrix} \right) - \text{rk}(\mathbf{H}_{12}) \nonumber \\
		&&\!\!\!\!\!\!\!\!\!\!\!\!\!\!\!\!\!\!\quad = \min \{N_1+M_2-D_{12},M_1+M_2\}.\label{EE04}
	\end{eqnarray}
\end{subequations}

For the second part, we divide \eqref{EE02} into two cases, and then expand them into the following.

For $N_2 \leq M_1$ Case, we have
\begin{subequations}
	\begin{eqnarray}
		&& \!\!\!\!\!\!\!\! \text{rk} \left( \begin{bmatrix} \mathbf{H}_{22} & \mathbf{H}_{21} \end{bmatrix} \right) = \min \{N_2,M_1+M_2\} = N_2, \\
		&& \!\!\!\!\!\!\!\! \text{rk} \left( \begin{bmatrix} \mathbf{H}_{11} \\ \mathbf{H}_{21} \end{bmatrix} \right) = \min \{M_1,D_{21}+N_1\}, \\
		&& \!\!\!\!\!\!\!\! \text{rk} \left( \begin{bmatrix} \mathbf{H}_{22} & \mathbf{H}_{21} \end{bmatrix} \right) + \text{rk} \left( \begin{bmatrix} \mathbf{H}_{11} \\ \mathbf{H}_{21} \end{bmatrix} \right) - \text{rk}(\mathbf{H}_{21}) \nonumber \\
		&&\!\!\!\!\!\!\!\! \quad = \min \{N_2+M_1-D_{21},N_1+N_2\}.\label{EE05}
	\end{eqnarray}
\end{subequations}

Otherwise, for $N_2 > M_1$ Case, we have
\begin{subequations}
	\begin{eqnarray}
		&&\!\!\!\!\!\!\!\!\!\!\!\!\!\!\!\!\text{rk} \left( \begin{bmatrix} \mathbf{H}_{22} & \mathbf{H}_{21} \end{bmatrix} \right) = \min \{N_2,M_2+D_{21}\}, \\
		&&\!\!\!\!\!\!\!\!\!\!\!\!\!\!\!\!\text{rk} \left( \begin{bmatrix} \mathbf{H}_{11} \\ \mathbf{H}_{21} \end{bmatrix} \right) = \min \{M_1,N_1+N_2\} = M_1, \\
		&&\!\!\!\!\!\!\!\!\!\!\!\!\!\!\!\!\text{rk} \left( \begin{bmatrix} \mathbf{H}_{22} & \mathbf{H}_{21} \end{bmatrix} \right) + \text{rk} \left( \begin{bmatrix} \mathbf{H}_{11} \\ \mathbf{H}_{21} \end{bmatrix} \right) - \text{rk}(\mathbf{H}_{21}) \nonumber \\
		&&\!\!\!\!\!\!\!\!\!\!\!\!\!\!\!\!\quad = \min \{N_2+M_1-D_{21},M_1+M_2\}.\label{EE06}
	\end{eqnarray}
\end{subequations}

Combining \eqref{EE03} and \eqref{EE04}, we can obtain 
\begin{equation}
	\begin{aligned}
		d_1 + d_2 &\leq \text{rk} \left( \begin{bmatrix} \mathbf{H}_{11} & \mathbf{H}_{12} \end{bmatrix} \right) + \text{rk} \left( \begin{bmatrix} \mathbf{H}_{12} \\ \mathbf{H}_{22} \end{bmatrix} \right) - \text{rk}(\mathbf{H}_{12}) \\  
		&\leq \min\{M_1 + M_2, N_1 + N_2, N_1 + M_2 - D_{12} \}.
	\end{aligned}
	\label{EE07}
\end{equation}

Combining \eqref{EE05} and  \eqref{EE06}, we can obtain 
\begin{equation}
	\begin{aligned}
		d_1 + d_2 &\leq \text{rk} \left( \begin{bmatrix} \mathbf{H}_{22} & \mathbf{H}_{21} \end{bmatrix} \right) + \text{rk} \left( \begin{bmatrix} \mathbf{H}_{11} \\ \mathbf{H}_{21} \end{bmatrix} \right) - \text{rk}(\mathbf{H}_{21}) \\
		&\leq \min\{M_1 + M_2, N_1 + N_2, N_2 + M_1 - D_{21} \}.
	\end{aligned}
	\label{EE08}
\end{equation}

Combining \eqref{EE07} and  \eqref{EE08}, we can obtain the desired upper bound 
\begin{equation}
	\begin{aligned}
		d_1 + d_2 &\leq \min\{M_1 + M_2, N_1 + N_2, \\
		&\quad M_1 + N_2 - D_{21}, M_2 + N_1 - D_{12}\}
	\end{aligned}
\end{equation}

\section{Conclusion}
In this paper, we established novel linear and numerical bounds for the active RIS-assisted MIMO wiretap IC, with linear beamforming techniques~\cite{Issa,Sridharan,Kao} and numerical bound derivation. Our contributions included: 1) we developed a linear and numerical lower bound through an enhanced transmission scheme building upon~\cite{Krishnamurthy} and~\cite{zheng2023dof}, formulated as an innovative linear integer programming problem for sum-SDoF lower bound maximization that  generalized~\cite{zheng2023dof} with security constraints; 2) we established a linear and numerical upper bound by combining the linear sum-DoF bound from~\cite[Theorem 4]{Tang} with nuclear norm minimization as a convex relaxation of the rank function; and 3) for symmetry antenna configurations, we obtained closed-form expressions for our lower bound through analysis. Extensive numerical simulations showed that our proposed bounds  tight alignment across many antenna configurations.

\begin{appendix}

\subsection{Proof of \eqref{8}}
To prove the relationship given by \eqref{8}, we take the first term $f_{\mathrm{e}1}\max\{N_\mathrm{e},M_1\}$ as an example and other terms can be derived in similar ways.
The linear equations given by \eqref{6a} when $i=1$ can be rewritten as
\begin{align}
	\mathbf{\Gamma}_{\mathrm{e},1}\boldsymbol{\phi} = \begin{bmatrix}
		\mathbf{0}_{ f_{e1}\max\{N_e,M_1\}}\\ 
		\mathbf{*}
	\end{bmatrix}+\tilde{\mathbf{h}}_{\mathrm{e}1},  
\end{align}
where $\tilde{\mathbf{h}}_{\mathrm{e}1}\triangleq\text{vec}(-\mathbf{H}_{\mathrm{e}1})$,  $\mathbf{\Gamma}_{\mathrm{e},1}\triangleq[\mathbf{V}_1^T,\mathbf{V}_2^T,...,\mathbf{V}_{N_\mathrm{e}}^T]$ with $[\mathbf{V}_k]_{i,j}=[\mathbf{G}_\mathrm{e}]_{k,j}[\mathbf{D}_1]_{j,i}$,  $\boldsymbol{\phi}\triangleq [\phi_1,\phi_2,...,\phi_R]^T$. It is observed that there are totally $f_{\mathrm{e}1}\max\{N_\mathrm{e},M_1\}$ linear equations. By summing all equations in \eqref{6a} and \eqref{6b}, we obtain $f_{e1} \max\{N_e, M_1\} + f_{e2} \max\{N_e, M_2\} +  \nonumber  f_{12}\max\{N_1,M_2\} + f_{21}\max\{N_2,M_1\}$ equations. To guarantee the solvability of this linear system (i.e., at least one solution), the number of equations must equal to or be less than the dimension of $\boldsymbol{\phi}$  (i.e., the number of variables) \cite{Gilbert}, which is $R$. Therefore, we can derive the relationship in \eqref{8}. 

\subsection{Proof of Tabular Closed-Form of Theorem 1 with Symmetry Antenna Configurations} 

For proving, we divide the antenna configurations into $4$ cases, and then maximize the sum-SDoF lower bound by balancing the constraints in Problem $(\text{P0}_1)$. 

For $N \le M \, \&\, N_e \le M$ Case, Problem $(\text{P0}_1)$ can be re-written as 
	\begin{subequations}
	\begin{eqnarray}
 \max_{				t, f_{e},  
				f \in \mathbb{Z}^+
		} && \!\!\!\!\!\!\!\! t \nonumber \\
		\text{s.t.} &&  \!\!\!\!\!\!\!\! t \le M - N_e + f + f_e, \label{30a} \\	 
		&& \!\!\!\!\!\!\!\! t \le 2\min\{N,M-N_{e} + f_e\},  \label{30b} \\		 
		&& \!\!\!\!\!\!\!\! f_e M + f M \le \frac{R}{2}.  \label{30c}
	\end{eqnarray}
\end{subequations} 
It can be seen that the optimal solution is given by balancing \eqref{30a} and \eqref{30b} and exhausting \eqref{30c}. That is
\begin{subequations}
	\begin{eqnarray}
	&&	\!\!\!\!\!\!\!\!\!\!\!\!\!\!\!\! M - N_e + f^* + f_e^* = 2\min\{N,M-N_{e} + f_e^*\}, \label{31a}\\
	&&	\!\!\!\!\!\!\!\!\!\!\!\!\!\!\!\!  f_e^* M + f^*M = \frac{R}{2}, \label{31b}
	\end{eqnarray}
\end{subequations}
where $f^*$, $f_e^* \in \mathbb{Z}^+$. By solving \eqref{31a}-\eqref{31b}, the  sum-SDoF lower bound is given by
\begin{equation}
	t^* = 2 \min \{M -N_e + \lfloor \frac{R - 2(M-N_e)M}{4M} \rfloor, N\}. 
\end{equation}

For $N \le M \, \&\, N_e > M$ Case, Problem $(\text{P0}_1)$ can be re-written as 
 	\begin{subequations}
 	\begin{eqnarray}
 		\max_{ 				t, f_{e}, 
 				f \in \mathbb{Z}^+
 		} && \!\!\!\!\!\!\!\! t \nonumber \\
 		\text{s.t.} &&  \!\!\!\!\!\!\!\! t \le f + f_e, \label{33a} \\	 
 		&& \!\!\!\!\!\!\!\! t \le 2\min\{N,f_e\},  \label{33b} \\		 
 		&& \!\!\!\!\!\!\!\! f_e N_e + f M \le \frac{R}{2}.  \label{33c}
 	\end{eqnarray}
 \end{subequations} 
It can be seen that the optimal solution is given by balancing \eqref{33a} and \eqref{33b} and exhausting \eqref{33c}. That is 
\begin{subequations}
	\begin{eqnarray}
		&&	\!\!\!\!\!\!\!\!\!\!\!\!\!\!\!\! f^* + f_e^* = 2\min\{N, f_e^*\}, \label{34a}\\
		&&	\!\!\!\!\!\!\!\!\!\!\!\!\!\!\!\!  f_e^* N_e + f^*M = \frac{R}{2}, \label{34b}
	\end{eqnarray}
\end{subequations}
where $f^*$, $f_e^* \in \mathbb{Z}^+$. By solving \eqref{34a}-\eqref{34b}, the  sum-SDoF lower bound is given by
\begin{equation}
	t^* = 2 \min \{\lfloor \frac{R}{2(N_e+M)} \rfloor, N\}. 
\end{equation}

For $N > M \, \& \, N_e \le M$ Case,  Problem $(\text{P0}_1)$ can be re-written as 
 	\begin{subequations}
	\begin{eqnarray}
		\max_{ 				t, f_{e},  
				f \in \mathbb{Z}^+
		} && \!\!\!\!\!\!\!\! t \nonumber \\
		\text{s.t.} &&  \!\!\!\!\!\!\!\! t \le N - N_e + f + f_e, \label{37a} \\	 
		&& \!\!\!\!\!\!\!\! t \le 2\min\{N, M - N_e + f_e\},  \label{37b} \\		 
		&& \!\!\!\!\!\!\!\! f_e M + f N \le \frac{R}{2}.  \label{37c}
	\end{eqnarray}
\end{subequations} 
It can be seen that the optimal solution is given by balancing \eqref{37a} and \eqref{37b} and exhausting \eqref{37c}. That is
\begin{subequations}
	\begin{eqnarray}
		&&	\!\!\!\!\!\!\!\!\!\!\!\!\!\!\!\! N - N_e + f^* + f_e^* = 2\min\{N, M - N_e + f_e^*\}, \label{38a}\\
		&&	\!\!\!\!\!\!\!\!\!\!\!\!\!\!\!\!  f_e^* M + f^* N = \frac{R}{2}, \label{38b}
	\end{eqnarray}
\end{subequations}
where $f^*$, $f_e^* \in \mathbb{Z}^+$. By solving \eqref{38a}-\eqref{38b}, the sum-SDoF lower bound is given by
 \begin{equation}
 	t^* = 2 \min \{M - N_e + \lfloor \frac{R - 2N(2M - N_e -N)}{2(M + N)} \rfloor, N\}. 
 \end{equation}

For $N > M \, \& \, N_e > M$ Case,  Problem $(\text{P0}_1)$ can be re-written as 
  	\begin{subequations}
 	\begin{eqnarray}
 		\max_{ 
 				t, f_{e}, 
 				f \in \mathbb{Z}^+
 		} && \!\!\!\!\!\!\!\! t \nonumber \\
 		\text{s.t.} &&  \!\!\!\!\!\!\!\! t \le N - M + f + f_e, \label{39a} \\	 
 		&& \!\!\!\!\!\!\!\! t \le 2\min\{N, f_e\},  \label{39b} \\		 
 		&& \!\!\!\!\!\!\!\! f_e N_e + f N \le \frac{R}{2}.  \label{39c}
 	\end{eqnarray}
 \end{subequations} 
 It can be seen that the optimal solution is given by balancing \eqref{39a} and \eqref{39b} and exhausting \eqref{39c}. That is
   	\begin{subequations}
 	\begin{eqnarray}
 			&&	\!\!\!\!\!\!\!\!\!\!\!\!\!\!\!\! N - M + f^* + f_e^* = 2\min\{N,   f_e^*\}, \label{40a}\\
 	&&	\!\!\!\!\!\!\!\!\!\!\!\!\!\!\!\!  f_e^* N_e + f^* N = \frac{R}{2}, \label{40b}	
  	\end{eqnarray}
\end{subequations} 		
where $f^*$, $f_e^* \in \mathbb{Z}^+$. By solving \eqref{40a}-\eqref{40b}, the  sum-SDoF lower bound is given by
\begin{equation}
	t^* = 2 \min \{\lfloor \frac{R - 2N(M-N)}{2(N+N_e)} \rfloor, N\}. 
\end{equation}

%\subsection{Proof of Tabular Closed-Form of Theorem 1 without Any Eavesdroppers} 
	
\end{appendix}

\bibliographystyle{IEEEtran}
    \bibliography{OJCOMS2025}

\end{document}